\documentclass[aps,prl,twocolumn,superscriptaddress]{revtex4-1} 
\usepackage{graphicx}
\usepackage{xcolor}
\usepackage{subfigure}
\usepackage{amsmath}
\usepackage{mathtools}
\usepackage{bm} 
\usepackage{xr}

\begin{document}

\title{Magnetic ground-state of the one-dimensional ferromagnetic chain compounds $M$(NCS)$_2$(thiourea)$_2$; $M$ = Ni, Co.}

\author{S. P. M. Curley}
\affiliation{Department of Physics, University of Warwick, Gibbert Hill Road, Coventry, CV4 7AL, UK}
\author{R. Scatena}
\affiliation{Department of Physics, Clarendon Laboratory, University of Oxford, Parks Rd., Oxford, OX1~3PU, United Kingdom}
\affiliation{Department of Chemistry and Biochemistry, University of Bern, 3012 Bern, Switzerland}
\author{R. C. Williams}
\author{P. A. Goddard}
    \email{p.goddard@warwick.ac.uk}
\affiliation{Department of Physics, University of Warwick, Gibbert Hill Road, Coventry, CV4 7AL, UK}
\author{P. Macchi}
\affiliation{Department of Chemistry, Materials, and Chemical Engineering, Polytechnic of Milan, Milan 20131, Italy}
\affiliation{Istituto Italiano di Tecnologia, Center for Nano Science and Technology CNST@polimi, Milan 20133, Italy}
\affiliation{Department of Chemistry and Biochemistry, University of Bern, 3012 Bern, Switzerland}
\author{T. J. Hicken}
\author{T. Lancaster}
\affiliation{Durham University, Department of Physics, South Road, Durham, DH1~3LE, United Kingdom}
\author{F. Xiao}
\affiliation{Laboratory for Neutron Scattering and Imaging, Paul Scherrer Institut, CH-5232 Villigen PSI, Switzerland}
\author{S. J. Blundell}
\affiliation{Department of Physics, Clarendon Laboratory, University of Oxford, Parks Road, Oxford, OX1~3PU, United Kingdom}
\author{V. Zapf}
\affiliation{National High Magnetic Field Laboratory, Materials Physics and Applications Division, Los Alamos National Laboratory, Los Alamos, New Mexico 87545, USA}
\author{J. C. Eckert}
\author{E. H. Krenkel}
\affiliation{Physics Department, Harvey Mudd College, Claremont, CA, United States}
\author{J. A. Villa}
\author{M. L. Rhodehouse}
\author{J. L. Manson}
    \email{jmanson@ewu.edu}
\affiliation{Department of Chemistry and Biochemistry, Eastern Washington University, Cheney, Washington 99004, USA}

\begin{abstract}

The magnetic properties of the two isostructural molecule-based magnets, Ni(NCS)$_{2}$(thiourea)$_{2}$, \textit{S} = 1, [thiourea = SC(NH$_2$)$_2$] and Co(NCS)$_{2}$(thiourea)$_{2}$, \textit{S} = 3/2, are characterised using several techniques in order to rationalise their relationship with structural parameters and ascertain magnetic changes caused by substitution of the spin. Zero-field heat capacity and muon-spin relaxation measurements reveal low-temperature long-range ordering in both compounds, in addition to Ising-like ($D < 0$) single-ion anisotropy ($D_{\rm{Co}} \sim$ -100 K, $D_{\rm{Ni}} \sim$ -10 K). Crystal and electronic structure, combined with DC-field magnetometry, affirm highly quasi-one-dimensional behaviour, with ferromagnetic intrachain exchange interactions \textit{J}$_{\rm{Co}}\approx~+4$ K and \textit{J}$_{\rm{Ni}}\sim~+100$ K and weak antiferromagnetic interchain exchange, on the order of $J'$ $\sim-~0.1$ K. Electron charge and spin-density mapping reveals through-space exchange as a mechanism to explain the large discrepancy in $J$-values despite, from a structural perspective, the highly similar exchange pathways in both materials. Both species can be compared to the similar compounds $M$Cl$_2$(thiourea)$_4$, $M$ = Ni(II) (DTN) and Co(II) (DTC), where DTN is known to harbour two magnetic field-induced quantum critical points. Direct comparison of DTN and DTC with the compounds studied here shows that substituting the halide Cl$^-$ ion, for the NCS$^-$ ion, results in a dramatic change in both the structural and magnetic properties.

\end{abstract}

\maketitle

\section{Introduction}

Constraining magnetic moments to lie and interact in one-dimensional chains or two-dimensional planes has, over the years, been an area of continued interest within the magnetism community \cite{Sachdev2000,Vasiliev2018,Giamarchi2008}. The reduced dimensionality generally serves to enhance quantum fluctuations, leading to the material hosting an array of exotic quantum ground states \cite{Haldane1983, Savary2016}. Several classes of low-dimensional materials can exhibit an order to disorder transition driven by an external tuning parameter, such as magnetic field, that pushes the system though a quantum critical point (QCP) \cite{Kosterlitz1973, Zapf2014}.

Considering the case of one-dimensional (1D) materials, a notable system is the ferromagnetically coupled effective $S$ = 1/2 chain material CoNb$_2$O$_6$. Applying a magnetic field transverse to Co(II) Ising-axis pushes the system through a QCP, as it moves from a ferromagnetic (FM) ordered ground-state to a disordered quantum paramagnetic phase \cite{Kinross2014,Coldea2010}. The existence of a QCP in 1D chains is not exclusive to the case of FM coupling. The $S$ = 1 antiferromagnetic (AFM) chain material NiCl$_2$(thiourea)$_4$ (DTN), where thiourea = SC(NH$_2$)$_2$, passes through two field-induced QCPs, at least one of which belongs to the universality class of a Bose-Einstein Condensate (BEC) \cite{Zapf2006a,Zvyagin2007}.


In terms of physically realising such systems, the use of organic ligands has proven highly effective in constructing crystal architectures that readily promote low-dimensional magnetic behaviour \cite{Brambleby2015,Hammar1999,Woodward,Goddard2008,Goddard2012}. A selection of recently published materials showcase the ability to test the limits of the theoretical understanding of $S$ = 1/2 and $S$ = 1 chain materials \cite{Liu2019, Huddart2019, Williams2020}. To achieve quasi-one-dimensional (Q1D) behaviour, the choice of intrachain bridging-ligand is a decisive one, as it ultimately determines the sign and strength of the intrachain exchange interaction ($J$). Non-bridging ligands also play an important role in promoting Q1D behaviour, keeping adjacent chains well separated and mitigating interactions due to interchain exchange ($J'$) \cite{Goddard2008,Goddard2012}. Characterising both $J$ and $J'$ is therefore paramount to establish how the crystal structure influences the magnetism in Q1D systems. The ultimate goal of this work is to move toward the construction of bespoke magnetic materials, where the magnetic properties can be chemically tuned from the point of synthesis. The work also provides an avenue to study how structural properties influence the observed ground-state, and possible emergence of quantum-critical behaviour, in low-dimensional magnetic systems.

\begin{figure}[t]
\centering
\includegraphics[width=1 \linewidth]{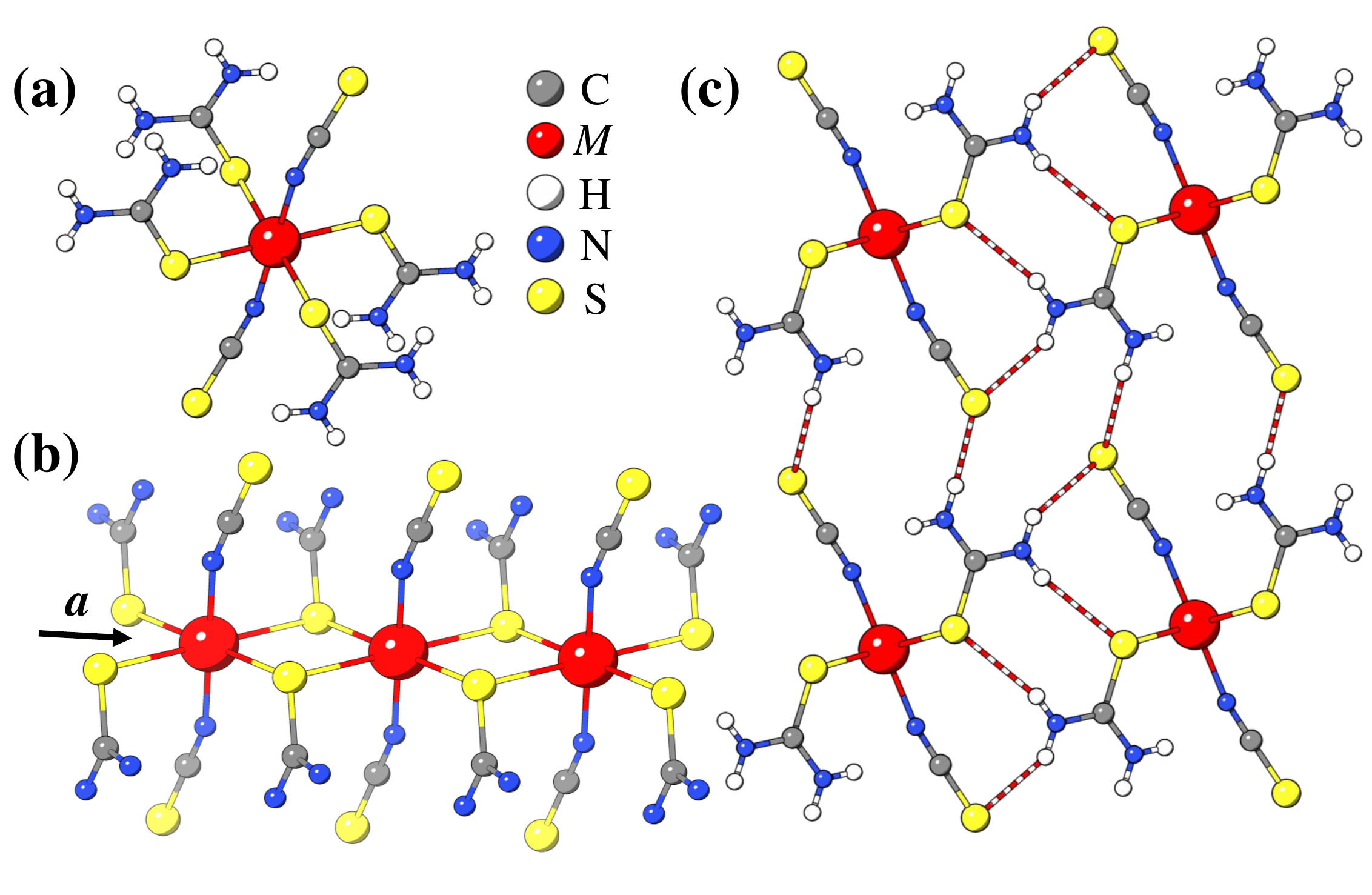}
\caption[width=1 \linewidth]{\small Structure of $M$(NCS)$_{2}$(thiourea)$_{2}$ where $M$ = Ni$^{2+}$, Co$^{2+}$. (a) Local $M^{2+}$ octahedral environment. (b) Two $M$---S---$M$ bonds form bibridge chains that propagate along the crystallographic $a$-axis. (c) H-bonding (striped-bonds) between adjacent chains within the \textit{bc}-plane. The structure is shown for $M$ = Co$^{2+}$ at $T$ = 100 K.}  \label{FigStruc}
\vspace{-0cm}
\end{figure}

We therefore turned our attention to the two isostructural coordination polymers Co(NCS)$_2$(thiourea)$_2$, $S$ = 3/2, and the $S$ = 1 analogue Ni(NCS)$_2$(thiourea)$_2$. Despite both compounds being first synthesised some time ago \cite{Yagupsky1965,Puglisi1967,Nardelli1966} we present here the first comprehensive study of their magnetic properties. The isostructural architecture of the two compounds, outlined below, permits us to investigate the effect the choice of transition metal ion has upon both the sign and strength of $J$ and $J'$, both of which govern the dimensionality of the system. An additional reason to study the properties of these materials is to establish their connection to known, chemically similar quantum magnets, such as the 1D chain DTN \cite{Figgis1986}. In the chosen materials the transition metal-ion ($M$) sits in a distorted $M$S$_{4}$N$_{2}$ octahedral environment, pictured in Figure~\ref{FigStruc}(a), suggesting the compounds likely possess a non-zero single-ion anisotropy (SIA) parameter ($D$) and rhombic anisotropy term ($E$). Along the crystallographic \textit{a}-axis, adjacent $M$-ions form chains through sulphide ions on two thiourea molecules forming two S-bridge pathways as shown in Figure~\ref{FigStruc}(b). The $M$---S---$M$ bond angles are close to 90$^{\circ}$, therefore, according to the Goodenough-Kanamori rules \cite{Goodenough1955, Kanamori1959}, these chain compounds represent promising platforms to investigate Q1D FM behaviour. The magnetic properties of both compounds within an applied magnetic field ($\mu_{\rm{0}}H$) can be summarised by the following Hamiltonian,

\noindent
\begin{multline}\label{eq:Hamiltonian}
\mathcal{H} = -J\sum\limits_{<i,j>} {\bf{\hat{S}}}_{i}\cdot{\bf{\hat{S}}}_{j} 
- J'\sum\limits_{<i,j'>}
{\bf{\hat{S}}}_{i}\cdot{\bf{\hat{S}}}_{j'}
+D\sum\limits_{i}
({{\bf{\hat{S}}}_{i}^{z}})^2 \\
+E\sum\limits_{i}
[({{\bf{\hat{S}}}_{i}^{x}})^2 - ({{\bf{\hat{S}}}_{i}^{y}})^2]
+ \mu_{\rm{B}}\mu_{0}\sum\limits_{i}\bf{H} \cdot \bf{g} \cdot \bf{\hat{S}}_{\textit{i}},
\end{multline}

\noindent
where $\bf{\hat{S}}_{\textit{i}}$ is the spin of each ion $i$, $<i,j>$ denotes a sum over nearest neighbours and a primed index in the summation describes the interaction with a nearest neighbour in an adjacent chain; $J>$ 0 corresponds to FM interactions. Here, $\bf{g} = \rm{diag}(g_x,g_y,g_z)$ is a tensor of $g$-factors where diagonal components are not necessarily equal. Whilst the triclinic structure of both materials are permissive of an $E$-term, the equatorial $M$S$_4$ environment (discussed in detail below) are only slightly distorted. As the departure from octahedral symmetry is limited, any $E$-term is expected to be small, as observed in similar Ni(II) and Co(II) complexes \cite{Ivanikova2006, Titis2007, Titis2011}.

Several methods were employed to investigate the electronic and magnetic properties of the two compounds in an attempt to ascertain the sign and magnitude of the dominant terms in Eq.~\ref{eq:Hamiltonian}: $J$, $J'$, and $D$. To identify likely magnetic exchange pathways, both crystal and electronic structures were inspected using multipolar model refinement of high resolution single-crystal X-ray diffraction data. Chemical bonding analysis was performed on the electron charge density distributions obtained from X-ray diffraction and density functional theory (DFT) calculations, and compared with calculated spin density maps to highlight the role of the magnetic ions. Heat capacity combined with muon-spin relaxation measurements revealed magnetic ordering and elucidated the magnitude of the SIA. DC-field magnetometry measurements, in combination with DFT calculations of the exchange coupling constants, helped illustrate a coherent picture of the magnetic ground state.


\section{Results}


\subsection{Crystal structure}

Crystallisation of Co(NCS)$_{2}$(thiourea)$_{2}$ and Ni(NCS)$_{2}$(thiourea)$_{2}$ coordination polymers resulted in needle-shaped single-crystals where single-crystal X-ray diffraction (SCXRD) measurements revealed twinning in both. Co(NCS)$_{2}$(thiourea)$_{2}$ could however also be obtained with thin-sheet morphology; these crystals showed clean single-crystal diffraction frames. Needle-like Ni(NCS)$_{2}$(thiourea)$_{2}$ crystals were irremediably affected by twinning and attempts to re-crystallise different morphologies proved unsuccessful. Both compounds crystallise into a triclinic structure with the space-group $P\overline{1}$, selected structural refinement parameters are listed in Table SIV of the supplemental information (SI) see \cite{SI} (as are the experimental details).

High quality Co(NCS)$_{2}$(thiourea)$_{2}$ single-crystals allowed X-ray diffraction data to be collected up to $d_{\rm{min}}=$ 0.50\,\AA, which, in combination with low-temperatures (\textit{T} = 100 K), allowed the refinement of the crystal structure, but also of the aspherical electron-charge density distribution, presented below. The Co atom resides on an inversion centre and lies in the middle of an octahedron built by pairs of Co---N [2.0226(4)~\AA], Co---S [2.5523(1)~\AA], and Co---S [2.5972(1)~\AA] coordination bonds. As a result of these bond lengths, the octahedral environment is axially compressed along the N---Co---N axis. The S---Co---S angles within the equatorial plane [84.19(1)$^{\circ}$ and 95.81(1)$^{\circ}$] and the angle between the octahedron axis and the equatorial plane [96.56(1)$^{\circ}$] show the Co(II) octahedra are slightly distorted. A Co---S---Co angle of 95.81(1)$^{\circ}$ defines the geometry of the two Co---S---Co bridges that make up the polymeric chain. The polymeric chains are packed together in the $bc$-plane by H-bonds in the range $3.43-3.47$\,\AA~and classify as weakly interacting.

Single-crystal X-ray diffraction data for Ni(NCS)$_{2}$(thiourea)$_{2}$ was measured at $T~=~173~$K up to $d_{\rm{min}}= 0.70$\,\AA~with a second twinning component  of $27.7(1)\%$ observed in the crystal. Analogous to the structure of Co(NCS)$_2$(thiourea)$_2$, the Ni atom sits on the inversion centre in the middle an octahedron built by two of each Ni---N [1.997(2)~\AA], Ni---S [2.5069(6)~\AA], and Ni---S [2.5517(6)~\AA] coordination bonds. The S---Ni---S angles within the equatorial plane [83.93(2)$^{\circ}$ and 96.07(2)$^{\circ}$] and the angle between the octahedron axis and the equatorial plane [97.20(7)$^{\circ}$] also show a slight distortion to the Ni(II) octahedra. A Ni---S---Ni angle of 96.07(2)$^{\circ}$ defines the geometry of two Ni---S---Ni bridges along the polymeric chain. H-bonds pack the chains in the $bc$-plane with distances between the donor and acceptor within the range $3.43-3.45$~\AA, which again classify as weakly bonded.

The experimentally-determined structural parameters of Co(NCS)$_{2}$(thiourea)$_{2}$ and Ni(NCS)$_{2}$(thiourea)$_{2}$ are very similar to one another. This similarity is found also from periodic DFT optimisation of the measured geometries, Table SVI \cite{SI}, which returned an even tighter correspondence between bond lengths and angles in the two species.

\subsection{Muon-spin relaxation}

\begin{figure}[t]
	\centering
	\includegraphics[width=\linewidth]{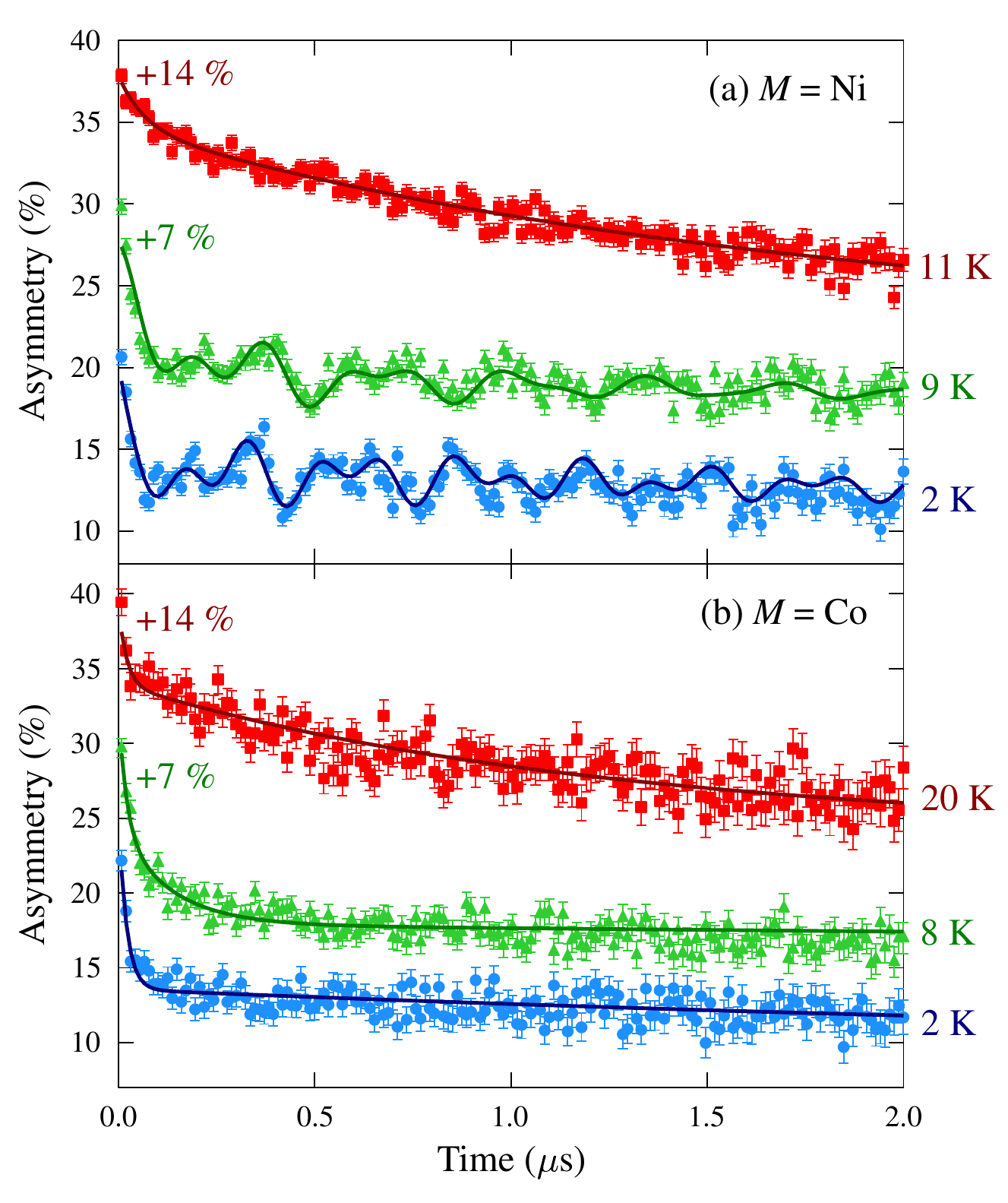}
	\caption{ZF $\mu^+$SR spectra measured on (a) Ni(NCS)$_2$(thiourea)$_2$ and (b) Co(NCS)$_2$(thiourea)$_2$. Data is presented with a vertical offset where needed for clarity. Fits shown as detailed in the Supplemental Material \cite{SI}}
	\label{fig:datfit}
\end{figure}

Zero-field positive-muon-spin-relaxation (ZF $\mu^+$SR) measurements on $M$(NCS)$_2$(thiourea)$_2$ were performed, with example spectra spanning the measured temperature range as shown in Figure~\ref{fig:datfit}.
The spectra measured for Co(NCS)$_2$(thiourea)$_2$ show no oscillations in the asymmetry, but
consist of two exponentially relaxing components, one with a large relaxation rate, found to be constant over the entire temperature range, and one with a smaller relaxation rate that dominates at later times; see SI  \cite{SI} and \cite{pratt2000wimda} for fit details.
The large relaxation rate is often observed in coordination polymers of this type and has been ascribed to a class of muon sites that are not well coupled to the magnetism (hence its temperature independence) that are realised close to electron density (e.g. double bonds, aromatic rings etc.) \cite{steele2011magnetic,Xiao2015}.

\begin{figure}[t]
	\centering
	\includegraphics[width=\linewidth]{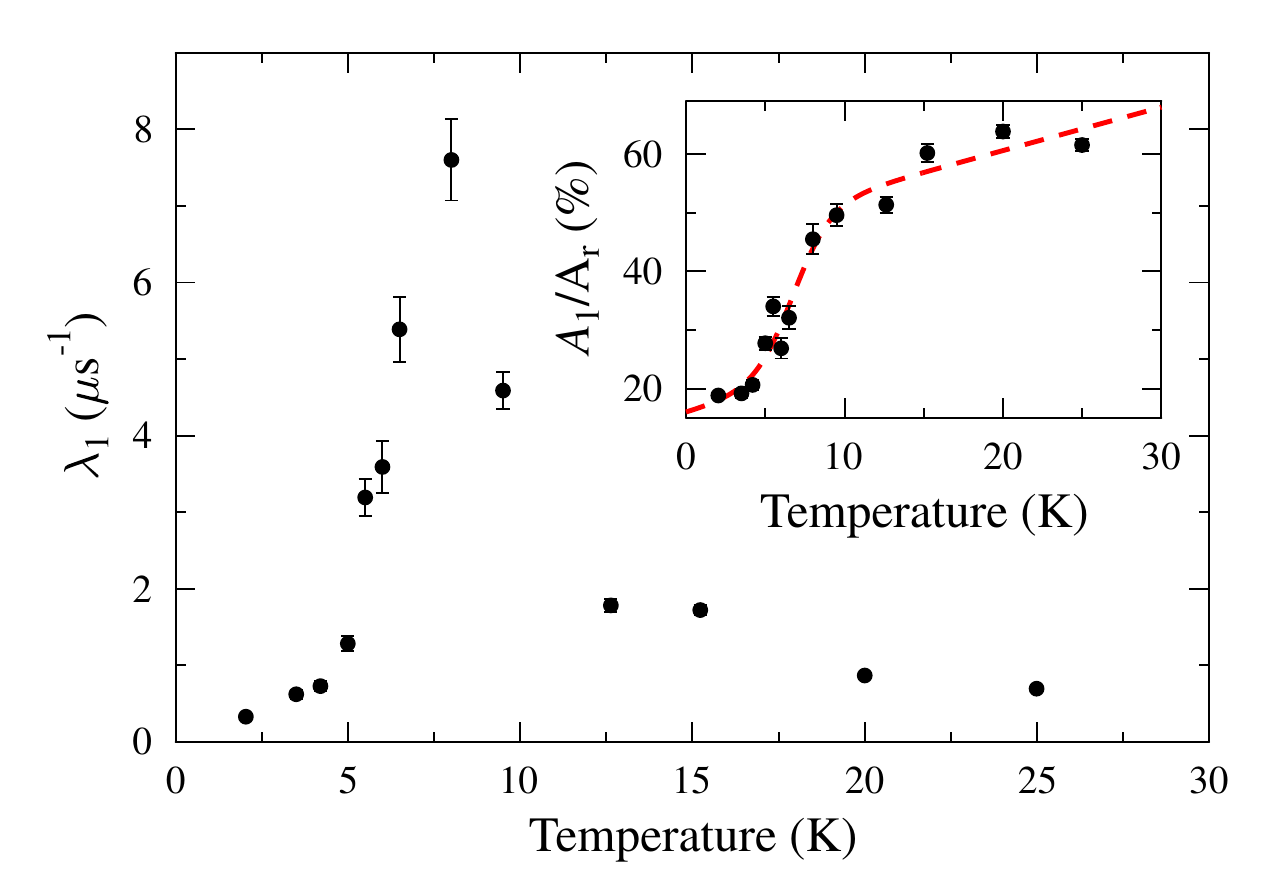}
	\caption{Temperature evolution of the small relaxation rate ($\lambda_1$) extracted through fitting ZF $\mu^+$SR asymmetry spectra measured on Co(NCS)$_2$(thiourea)$_2$. Inset shows the temperature evolution of the ratio of the amplitude of the component with small relaxation rate ($A_{1}$) to the total relaxing asymmetry ($A_\textrm{r}$). Line is a guide to the eye.}
	\label{fig:Co_lambdaA}
\end{figure}

The temperature evolution of the smaller relaxation rate ($\lambda_{1}$) is shown in Figure~\ref{fig:Co_lambdaA}.
It exhibits a peak around 8~K, indicating a phase transition from a magnetically ordered to disordered state, in good agreement with heat-capacity and magnetometry measurements discussed below.
The lack of oscillations at low temperature and the observed exponential relaxation suggest that the system is dominated by dynamic fluctuations on the muon timescale, such that coherent precession of the muon-spin is not measured. This has been noted in several coordination polymer magnets containing Fe$^{2+}$ \cite{lancaster2006muon}, Mn$^{2+}$ \cite{manson2013mn} and (in some cases) Ni$^{2+}$ \cite{liu2016}, where the sizeable magnetic moment can lead to a large, fluctuating distribution of local magnetic fields at the muon sites. (In contrast, coordination polymer magnets containing Cu$^{2+}$ often show oscillations in the ordered regime~\cite{steele2011magnetic}.)
The observed exponential relaxation suggests that the sample is in the fast-fluctuation regime below $T_\text{c}$, where the relaxation rate varies as $\lambda\propto \Delta^{2}\tau$.
Here $\Delta$ is the variance of the field distribution sampled by the muons and $\tau$ is the fluctuation time. 
The variance $\Delta$ varies with the size of the local magnetic field, so we might expect materials with larger moments to lead to larger relaxation rates. In this scenario the relaxation rate will scale faster with moment than the oscillation frequency, and therefore a higher moment could prevent the observation of coherent precession.
This could explain the observed lack of oscillations in this system compared to similar materials with smaller moments.
It is also possible that the lack of oscillations reflects a greater propensity for Co-based systems to adopt magnetic structures that yield inhomogeneous local field distributions or that give dynamic fluctuations in the muon (MHz) timeframe.

Further evidence for a transition around 8~K comes from the rapid change in amplitude of the small relaxation rate component ($A_1$), as shown in the inset of Figure~\ref{fig:Co_lambdaA}. (This effect has also been observed in similar materials close to the transition temperature \cite{lancaster2006muon, manson2013mn,liu2016}.)  Significant in this case is that both this change and the peak in $\lambda$ persist over a relatively wide range of temperature, extending at least 2--3~K below $T_\text{c}$. This suggests that the phase transition is extended in temperature on the MHz timescale with an onset of dynamics occurring in the ordered regime above 5~K.

In contrast to the Co material, ZF $\mu^+$SR spectra for Ni(NCS)$_2$(thiourea)$_2$ show oscillations in the asymmetry for temperatures $T \lesssim~10.4$~K, indicating coherent muon-spin precession consistent with long-range magnetic order.
Two oscillation frequencies [in constant ratio $\nu_2/\nu_1~=~1.754(1)$] were observable over the whole temperature range, with a third $\nu_3~\simeq~3\nu_1$ only observable for $T~\lesssim~7$~K.
This indicates three distinct muon stopping sites in this material.

\begin{figure}[t]
	\centering
	\includegraphics[width=\linewidth]{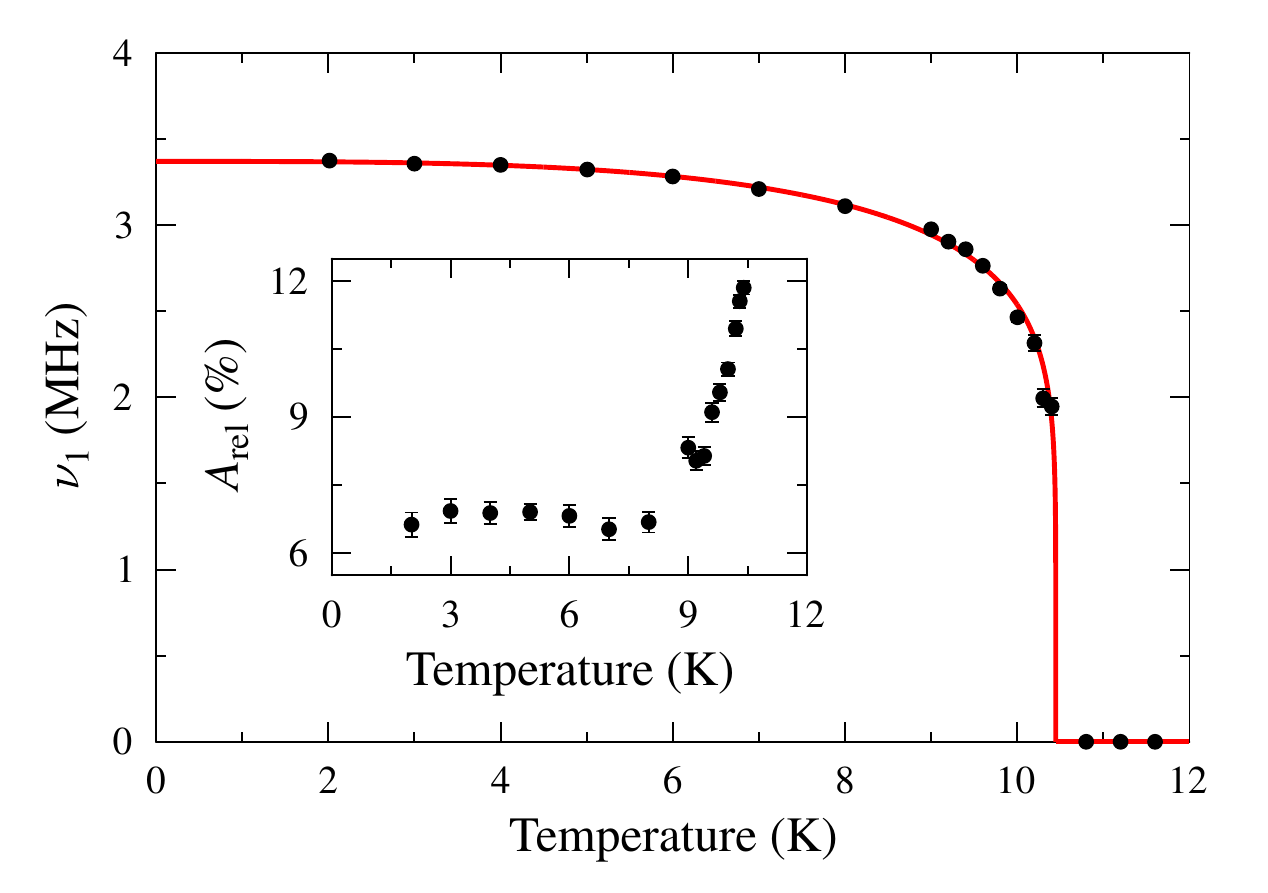}
	\caption{Temperature evolution of the smallest oscillation frequency ($\nu_1$) observable in the ZF $\mu^+$SR asymmetry spectra measured on Ni(NCS)$_2$(thiourea)$_2$, with a fit to Eqn.~\ref{eqn:crit_bhv}. Inset shows the temperature evolution of the total relaxing asymmetry ($A_{\rm{rel}}$) from the same measurements.}
	\label{fig:Ni_freqrel}
\end{figure}

By fitting these spectra as seen in Figure~\ref{fig:datfit}(b), see SI~\cite{SI} for fit details, the frequencies were extracted.
The smallest frequency was fitted to
\noindent
\begin{equation}\label{eqn:crit_bhv}
	\nu_i(T) = \nu_i(0)\left[1-\left(\frac{T}{T_\text{N}}\right)^\delta\right]^\beta,
\end{equation}
\noindent
as seen in Figure~\ref{fig:Ni_freqrel}.
It was found that $\nu_1(0)~=~3.37(2)$~MHz [corresponding to a local magnetic field of 24.1(1)~mT at the muon-spin site at zero-temperature], $\delta~=~3.2(3)$, $\beta~=~0.14(3)$ and $T_\text{N}~=~10.4(1)$~K.
This value of $\beta$ is consistent with a system dominated by 2D magnetic fluctuations, being very close to the value for the 2D Ising model, $\beta~=~1/8$ \cite{binney1992theory}.
This transition temperature is also supported by the behaviour of relaxation rates of the oscillating components, which both show a peak between $T~=10.3$~K and $T~=10.4$~K.
These rates diverge when approached from below, indicating critical slowing down of magnetic fluctuations, which often occurs in the proximity of a magnetic transition.

The disappearance of the third frequency, in combination with the increase in the relaxing asymmetry ($A_{\rm{rel}}$) for $T~\gtrsim~7$~K, seen in the inset of Figure~\ref{fig:Ni_freqrel}, is evidence for the onset of the magnetic transition in this material, suggesting again, a broad phase transition taking place in the region 7~$\lesssim~T~\lesssim$~10.4~K on the muon timescale.
For $T~\gtrsim~10.4$~K no oscillations are observable in the asymmetry spectra, consistent with the system being magnetically disordered.



\subsection{Heat capacity}

\begin{figure}[t]
\centering
\includegraphics[width=\linewidth]{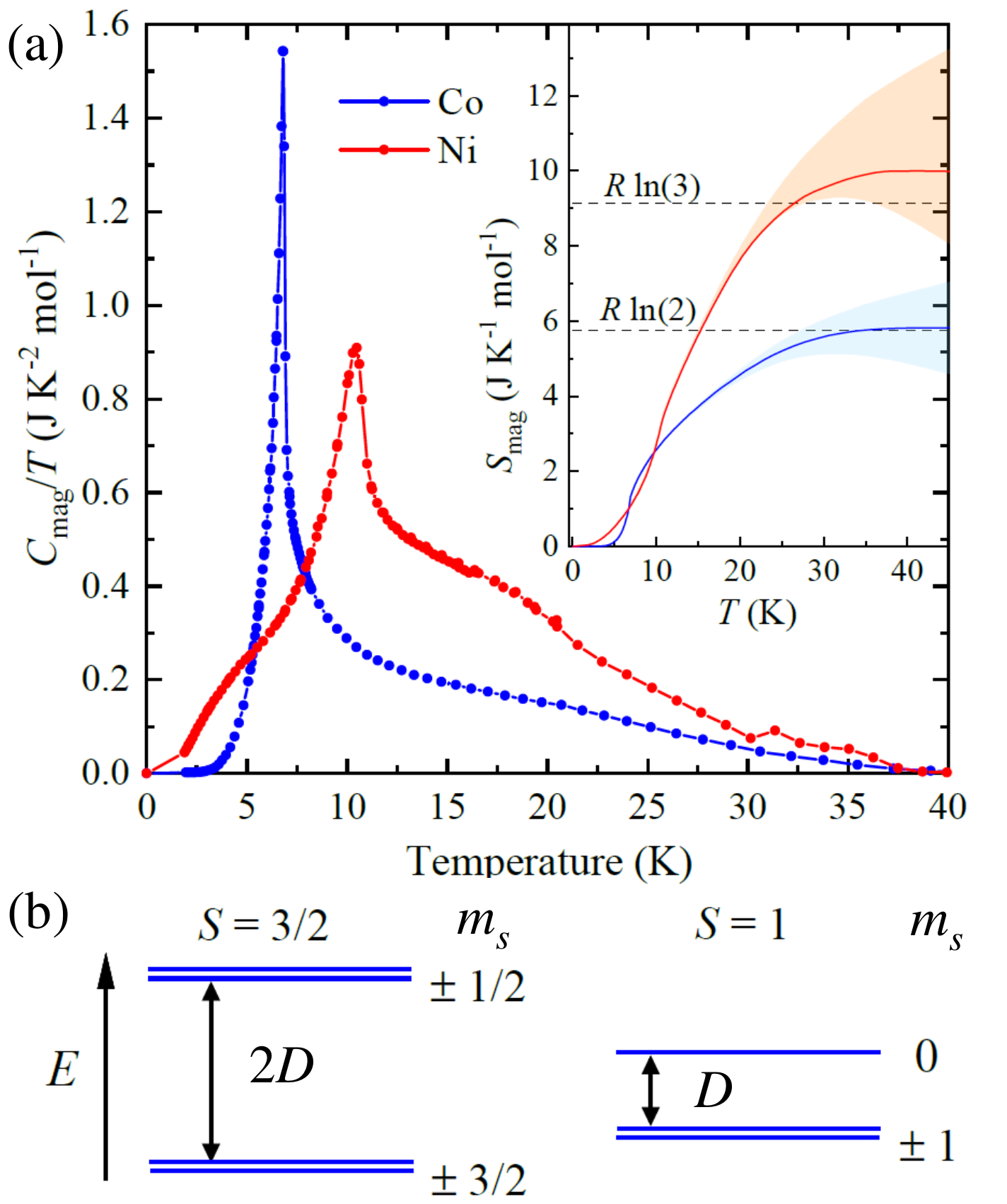}
\caption{\small (a) Zero-field magnetic heat capacity ($C_{\rm{mag}}$) divided by temperature ($T$) and plotted as a function of $T$. Inset shows the resultant magnetic entropy ($S_{\rm{mag}}$), where the shaded regions are representative of errors introduced by the uncertainty in the amplitudes of the high-temperature fit components. (b) Energy ($E$) level diagrams for $S$ = 3/2 and $S$ = 1 moments within octahedral environments with easy-axis single-ion anisotropy ($D < 0$) where $m_s$ is the eigenvalue of the spin operator $S_z$.} \label{HC}
\vspace{-0cm}
\end{figure}

Zero-field heat capacity ($C$) was measured as a function of temperature for a single-crystal of Co(NCS)$_{2}$(thiourea)$_{2}$ and a polycrystalline pressed pellet of Ni(NCS)$_{2}$(thiourea)$_{2}$. In order to isolate the low-temperature magnetic heat capacity ($C_{\rm{mag}}$) the high-temperature ($T \gtrsim 30$~K) contribution was reproduced using a phenomenological model containing both Debye and Einstein phonon modes and subtracted as a background \cite{SI,Lashley2003}. For both compounds, a $\lambda$-peak in $C_{\rm{mag}}$ at low-temperatures, seen in Figure \ref{HC}(a), is indicative of a transition to a magnetically-ordered state, giving ordering temperatures of $T_{c}$ = 6.82(5) K and 10.5(1) K for the Co and Ni species, respectively, in excellent agreement with ZF $\mu^+$SR.

Co(NCS)$_{2}$(thiourea)$_{2}$ $C_{\rm{mag}}$ data possesses a notably sharper $\lambda$-peak compared to Ni(NCS)$_{2}$(thiourea)$_{2}$. Single-crystal magnetometry for the Co species, discussed below, provides the reason for this. The magnetometry data indicate an AFM ground-state, where the spins possess a strong Ising-like SIA. For a Q1D chain of Ising spins, a broad-hump in $C_{\rm{mag}}$ is expected as the reduced dimensionality causes the build up of spin-spin correlations at temperatures above the transition temperature \cite{Honmura1984,Feng2010}. The sharp nature of the peak points to 2D or 3D Ising-like long range ordering within the material. The measured response of $C_{\rm{mag}}(T)$ was not well captured by simulations of $C_{\rm{mag}}(T)$ for the 2D Ising model \cite{Onsager1944,McCoy1973}; Figure S2 \cite{SI}. This would imply the secondary exchange along $b$ and $c$ are similar in magnitude, which is also supported by the DFT calculations (below) and indicates $n' = 4$ next nearest neighbours.

The magnetic entropy per mole ($S_{\rm{mag}}$) for both compounds can be determined using,

\noindent
\begin{equation}
S_{\rm mag}(T) = \int_0^{T}  {\rm d}T'~ \frac{C_{\rm mag}(T')}{T'},
\end{equation}
\noindent

\noindent
where it is assumed $C_{\rm{mag}}$ = 0 at $T$ = 0 K. The results are shown in the inset of Figure \ref{HC}(a). For Ni(NCS)$_{2}$(thiourea)$_{2}$, $S_{\rm{mag}}$ fully recovers to $R$ln(3) ($R$ is the ideal gas constant) as expected for a $S$ = 1 ion. In contrast, Co(NCS)$_{2}$(thiourea)$_{2}$ shows an initial sharp upturn before a broad rise to a plateau at $R$ln(2) around 30 K. Electronic structure calculations reveal the Co ion sits in the high-spin Co$^{2+}$ state. The spin configuration consists of two Kramer-doublets split by an energy-gap of magnitude 2$D$ [see Figure \ref{HC}(b)]. For a \textit{S} = 3/2 ion, we expect $S_{\rm{mag}}$ to recover to \textit{R}ln(4) in accordance with \textit{R}ln(2\textit{S}+1). A recovery to \textit{R}ln(2) is indicative of a system exhibiting large SIA which keeps the two Kramer-doublets well separated as illustrated in Figure \ref{HC}(b), suggesting that the system can be well approximated using an effective spin-half approach within the low-temperature regime. We therefore expect $D \sim - 100~{\rm K}$ as seen in similar easy-axis Co(II) complexes \cite{Titis2011}.


\begin{figure}[t]
\centering
\includegraphics[width=\linewidth]{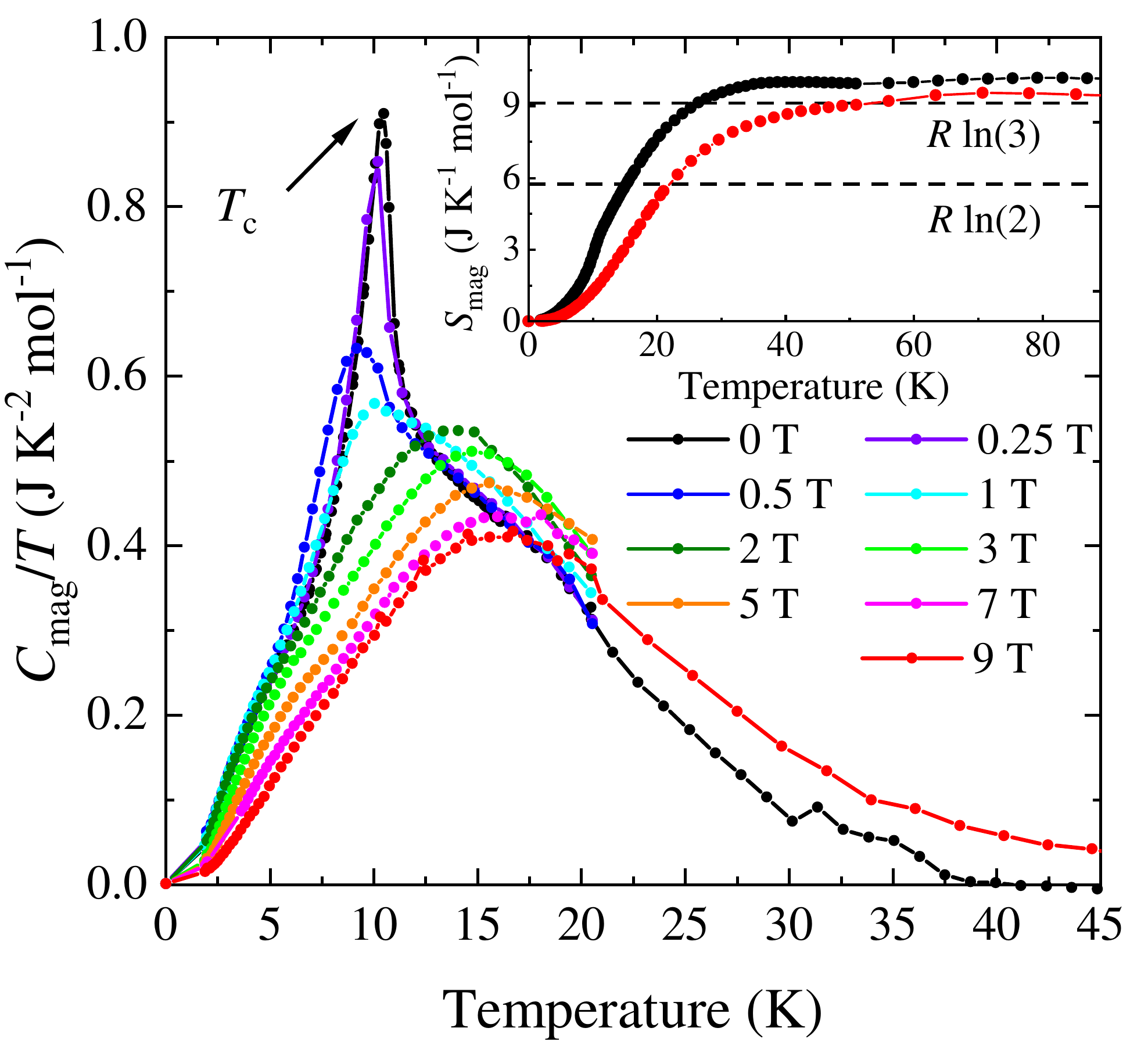} 
\caption{\small Magnetic heat capacity ($C_{\rm{mag}}$) divided by temperature ($T$) and plotted as a function of $T$ for a polycrystalline pressed pellet of Ni(NCS)$_{2}$(thiourea)$_{2}$ measured in applied fields up to 9 T. The zero-field transition temperature to a magnetically ordered state is indicated at $T_{c}$. Inset shows magnetic entropy ($S_{\rm{mag}}$) recover to $R$ln(3) at 0 and 9 T, as expected for a \textit{S} = 1 ion.}  
\label{fig:Ni_HC}
\end{figure}
\vspace{-0cm}

The behaviour of the Ni-species was investigated further by measuring $C_{\rm{mag}}$ in applied magnetic fields up to 9 T as shown in Figure \ref{fig:Ni_HC}. The $\lambda$-peak is initially suppressed in field, suggesting the ground-state is AFM in nature. At higher fields, the ordering peak ceases to be resolvable and the broad hump shifts to higher temperatures reflecting the Schottky-like response to the field induced Zeeman splitting of the ground-state energy levels. The inset of Figure 5 shows $S_{\rm{mag}}$ fully recovers to $R$ln(3) on warming above about 40 K in zero field and 50 K at 9 T, indicating the level splitting is completely overcome by these temperatures. This is consistent with $|D| \sim 10$ K, which is typical for Ni$^{2+}$ ions in similar local environments \cite{Baran2005b,Titis2010}.

\subsection{Magnetometry}




Single-crystal magnetometry measurements were performed on Co(NCS)$_{2}$(thiourea)$_{2}$ with the external magnetic-field applied along three orthogonal orientations: parallel to the chain axis $a$, perpendicular to $a$ within the $ab$-plane denoted $b'$ and normal to the $ab$-plane denoted $c'$. The angle between the magnetic field and the unique N---Co---N axis is 84$^{\circ}$ for $\mu_{\rm{0}}H$ $\parallel a$, 73$^{\circ}$ for $\mu_{\rm{0}}H$ $\parallel$ $b'$ and 18$^{\circ}$ for $\mu_{\rm{0}}H$ $\parallel c'$. 

\begin{figure}[t]
\centering
\includegraphics[width= 1 \linewidth]{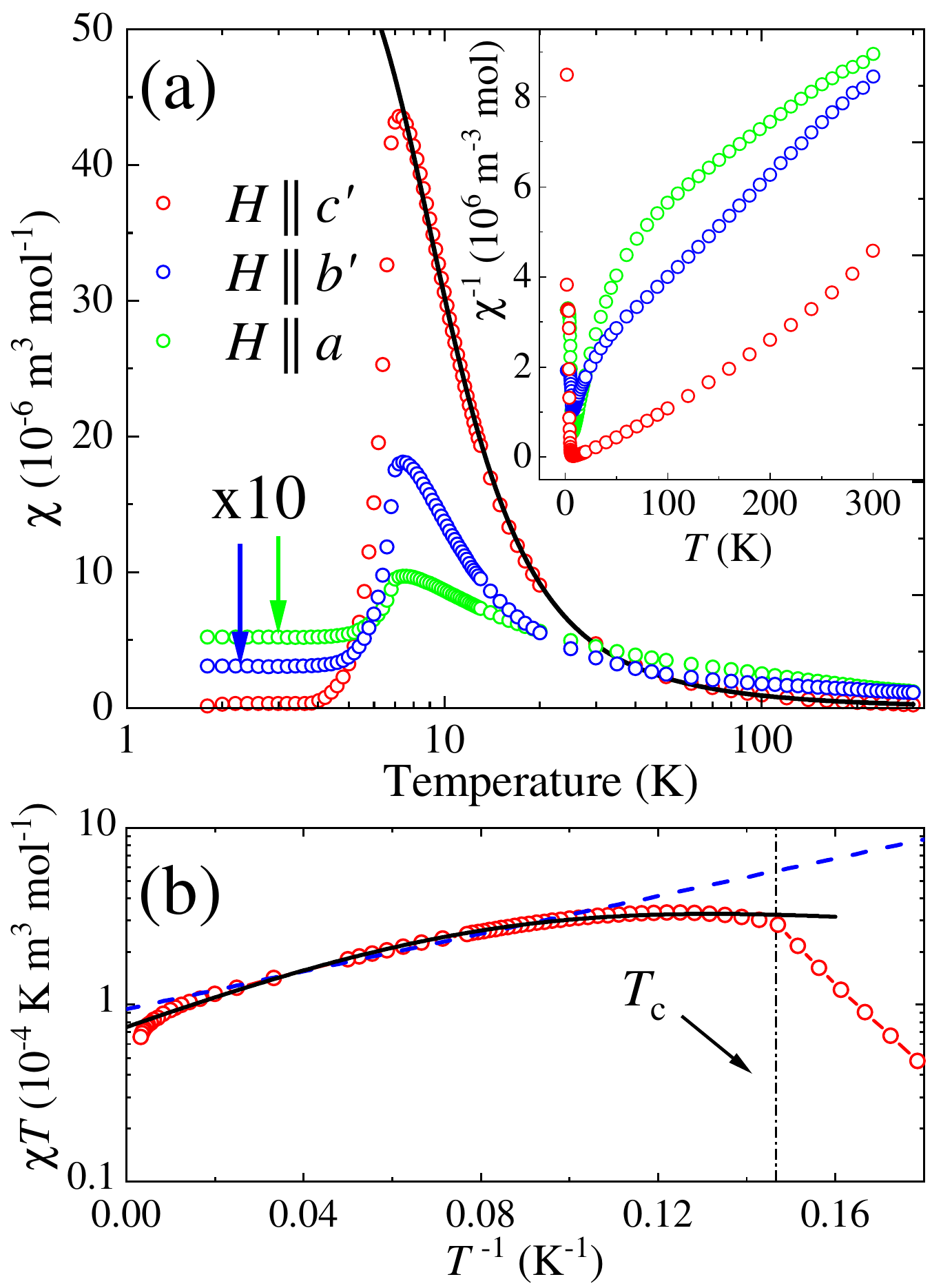}
\caption{Single-crystal magnetic susceptibility $\chi(T)$ data for Co(NCS)$_{2}$(thiourea)$_{2}$ measured at $\mu_{0}H$ = 0.1 T. \small(a) $\chi(T)$ for different field directions where data for $H \parallel b'$ and $a$ are multiplied by a factor of 10. Inset shows a plot of $\chi^{-1}(T)$ with no scaling. \small(b) Semi-logarithmic plot of $\chi T(T)$ against $T^{-1}$ (discussed in text) for field parallel to $c'$. Dashed line is a fit to $\chi^{||}_{1\rm{D}}$ (see text). Solid line in both panels is a fit to Eq.~\ref{eq:chi_MF} within the temperature range 10 K $\leq T \leq$ 100 K.}  \label{fig:Co_MvT}
\vspace{-0cm}
\end{figure}

Figure \ref{fig:Co_MvT}(a) shows that upon decreasing temperature, the magnetic susceptibility [$\chi$($T$)] for field along all three orientations rises to a sharp peak before plateauing at low-temperatures, behaviour indicative of an AFM ground state. The Fisher relation \cite{Fisher1962} estimates a transition temperature $T_{c}$ = 6.7(1) K, in excellent agreement with heat capacity and $\mu^{+}$SR data. The $\chi(T)$ with field parallel to $c'$, red circles Figure \ref{fig:Co_MvT}(a), is an order of magnitude greater than measurements made with the field along $a$ or $b'$, verifying a strong Ising-like ($D < 0$) SIA. Due to the close proximity of $c'$ to the unique axial N---Co---N bond, it is highly likely that the magnetic moments on the Co ions are co-linear with the N---Co---N axis.

\noindent
The $\chi^{-1}(T)$ data for all three field directions, inset Figure \ref{fig:Co_MvT}(a), shows curvature persisting up to the highest measured temperature ($T = 300~\rm{K}$). This non-Curie-like behaviour suggests that the leading energy term in the Hamiltonian is similar in size to the thermal energy in this temperature range. On the basis of our heat capacity measurements, we expect this term to be $D$. Thus we estimate $|D|\sim 100$~K in agreement with heat capacity data and similar Co(II) complexes \cite{Titis2011}.

The large negative SIA (zero-field splitting = $2|D|\sim 200$~K) suggests that as temperature is lowered below 100~K, a vanishingly small proportion of the spins will populate the excited doublet and the magnetic properties can be accounted for within an effective spin-half ($S_{\rm eff} = 1/2$) Ising model. This means that over the temperature range $10 \leq T \leq 100$~K, the susceptibility for $H \parallel c'$ (which is close to parallel to the expected Ising axis) can be approximated by that of the ideal 1D $S$ = 1/2 Ising chain [$\chi^{||}_{1\rm{D}}(T)$] and parameterised by $J_{\rm eff}$, the primary exchange energy in the effective $S_{\rm eff}$ = 1/2 picture \cite{Fisher1963,Greeney1989}. Deviations from strictly 1D behaviour can be accounted for by introducing a mean-field correction to the susceptibility given by the expression \cite{Greeney1989},

\noindent
\begin{equation}\label{eq:chi_MF}
\chi_{\rm{mf}} = \frac{\chi^{||}_{1\rm{D}}(T)}{[1-(n'J'_{\rm{eff}}) \chi^{||}_{1\rm{D}}(T)/C_{\rm{1D}}]},
\end{equation}

\noindent
where $C_{\rm{1D}}$ is the easy-axis Curie constant for the idealised 1D picture and $J'_{\rm{eff}}$ is the interchain exchange in the effective $S_{\rm{eff}}$ = 1/2 picture arising from $n'$ interchain nearest neighbours where $n'= 4$ for this material. 
  
Following the analysis by Greeney \textit{et al.} \cite{Greeney1989} in Figure \ref{fig:Co_MvT}(b) we show a semi-logarithmic plot of $\chi T(T)$ against $T^{-1}$ for $H \parallel c'$. In this diagram $\chi^{||}_{1\rm{D}}(T)$ is a straight line with gradient given by the intrachain exchange $J_{\rm{eff}}$ and intercept related to $C_{\rm{1D}}$. A positive slope is indicative of FM intrachain exchange ($J_{\rm{eff}} >$ 0). The sharp kink at $T_{\rm{c}}$ indicates the onset of long-range order. The data near $T_{\rm{c}}$ deviates from the linear response predicted by the ideal 1D Ising model (dashed line). A fit to Eq.~\ref{eq:chi_MF} within the temperature range $10~\rm{K} \leq T \leq 100~\rm{K}$ ($T < 2|D|$, well within the $S_{\rm{eff}} = 1/2$ regime) is found to more accurately track the data [solid line, Figure \ref{fig:Co_MvT} (a-b)], indicating the importance of the intrachain exchange interactions. The fit to Eq.~\ref{eq:chi_MF} returned parameters of $g_{\rm{eff}} = 8.0(1)$, $J_{\rm{eff}} = 10.4(2)~{\rm K}$ and $J'_{\rm{eff}}=-0.31(2)~{\rm K}$. The fit parameters are consistent with the effective $S$ = 1/2 model, in which the full Co(II) moment of the ground state doublet is absorbed into the effective $g$-factor and exchange energies. The real exchange, $J$, is related to the effective value via $J = (3/5)^{2}J_{\rm{eff}}$ \cite{Lines1971, Lloret2008}, hence the values we extract are $J = 3.62(7)~\rm{K}$ and $J' = -0.12(1)~\rm{K}$. The fit to Eq. \ref{eq:chi_MF} deviates from the data near the onset of long-range order. The departure at high temperatures is caused by the breakdown of the Ising model as temperatures approach $|D| \sim 100$ K. The difference between $\chi(T)$ for the field along $b'$ and $a$ may be due to either, the existence of a small $E$-term, a slight misalignment of the magnetic field or different temperature dependencies of the effective magnetic moment along $b'$ and $a$.



\begin{figure}[t]
\centering
\includegraphics[width=1\linewidth]{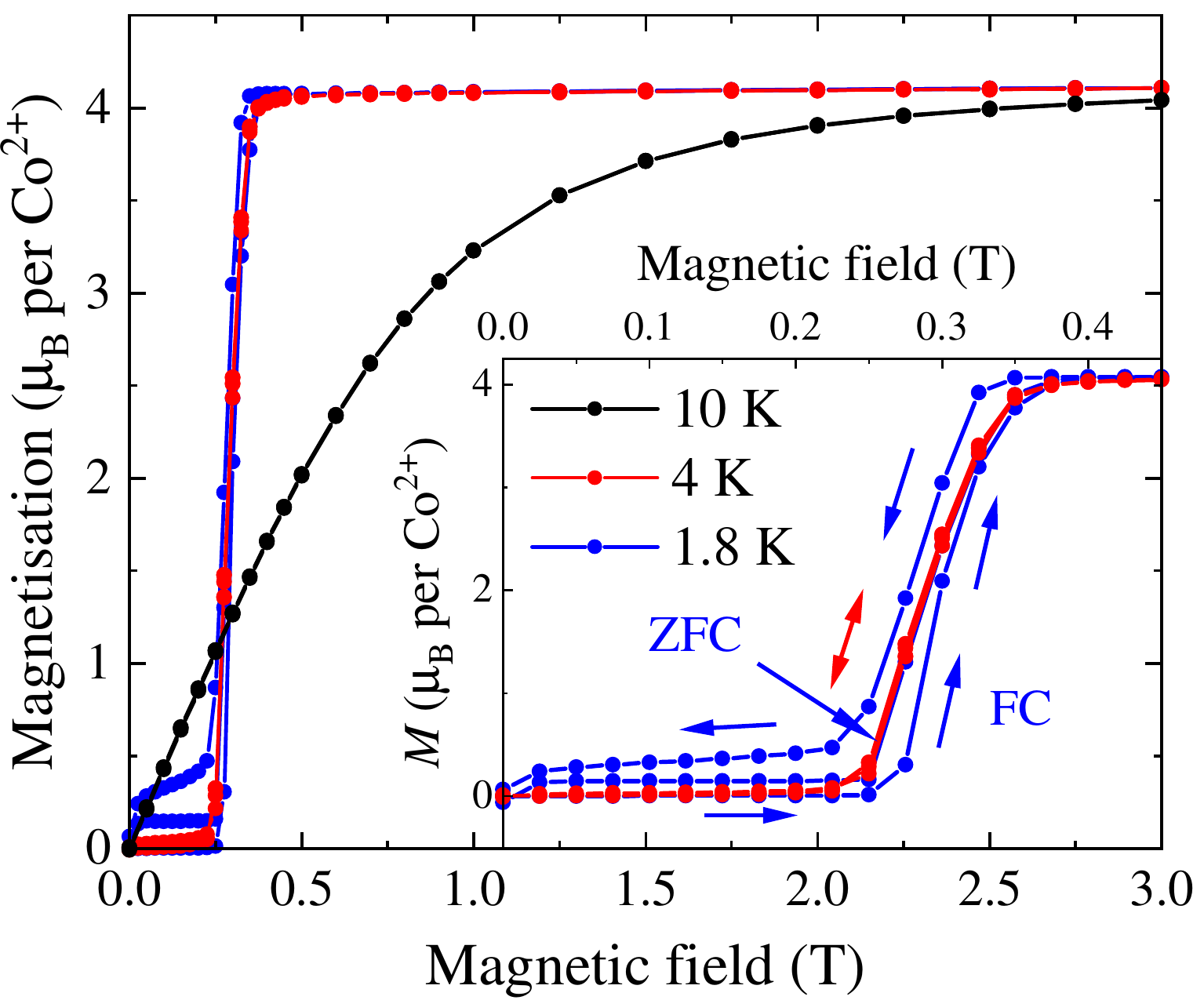}
\caption{\small Single-crystal DC-field magnetisation $M$($H$) data for Co(NCS)$_{2}$(thiourea)$_{2}$ with the field parallel to $c'$. Inset highlights the low-field hysteretic behaviour seen at the lowest temperature. Zero-field cooled (ZFC) and field cooled (FC) up-sweeps are indicated with arrows.}  \label{fig:Co_MvH}
\vspace{-0cm}
\end{figure}
\noindent

The magnetisation $M$($H$) of Co(NCS)$_2$(thiourea)$_2$ with field parallel to $c'$ is shown in Figure \ref{fig:Co_MvH}. At $T \leq 4$~K, the induced moment is approximately zero at low fields before rapidly rising to a saturation moment of $M_{\rm{sat}} = 4.1(1)~\mu_{\rm{B}}$ per Co$^{2+}$ ion at $\mu_{\rm{0}}H_{\rm{sf}} = 0.29(5)~$\,T. We attribute this feature to spin-flip behaviour where the spins are rapidly polarised from their AFM ground-state to an FM saturated state as the interchain AFM bonds are overcome by the Zeeman interaction. The step broadens and disappears at $T > 4~$K.

Within the $S_{\rm{eff}} = 1/2$ model $M_{\rm{sat}}$ yields $ g_{\rm{eff}} = 8.2(1)$, in excellent agreement with the result from fitting $\chi(T)$. The field at which the spin-flip occurs can be related to the AFM interaction strength via $ g_{\rm{eff}} \mu_{\rm B} \mu_{\rm{0}}H_{\rm{sf}} = S n' J'_{\rm{eff}}$. Assuming $n' = 4$, we obtain $J' = -0.3(1)~\rm{K}$ which is in reasonable agreement with the value extracted from $\chi(T)$. As $M$($H$) directly probes the AFM interchain bonds at low-temperature, during the spin-reversal process, we expect $M$($H$) to provide us with the more trustworthy estimate of $J'$. The inset to Figure~\ref{fig:Co_MvH}~ shows hysteresis in $M$($H$) for $T < 4~\rm{K}$, expected for FM coupled Ising spins. This behaviour is also observed in the similar FM Ising chain compound Co(NCS)$_2$(4-benzoylpyridine)$_2$ (4-benzoylpyridine = C$_{12}$H$_9$NO) \cite{Rams2017}.

\begin{figure}[t!]
\centering
\includegraphics[width=1\linewidth]{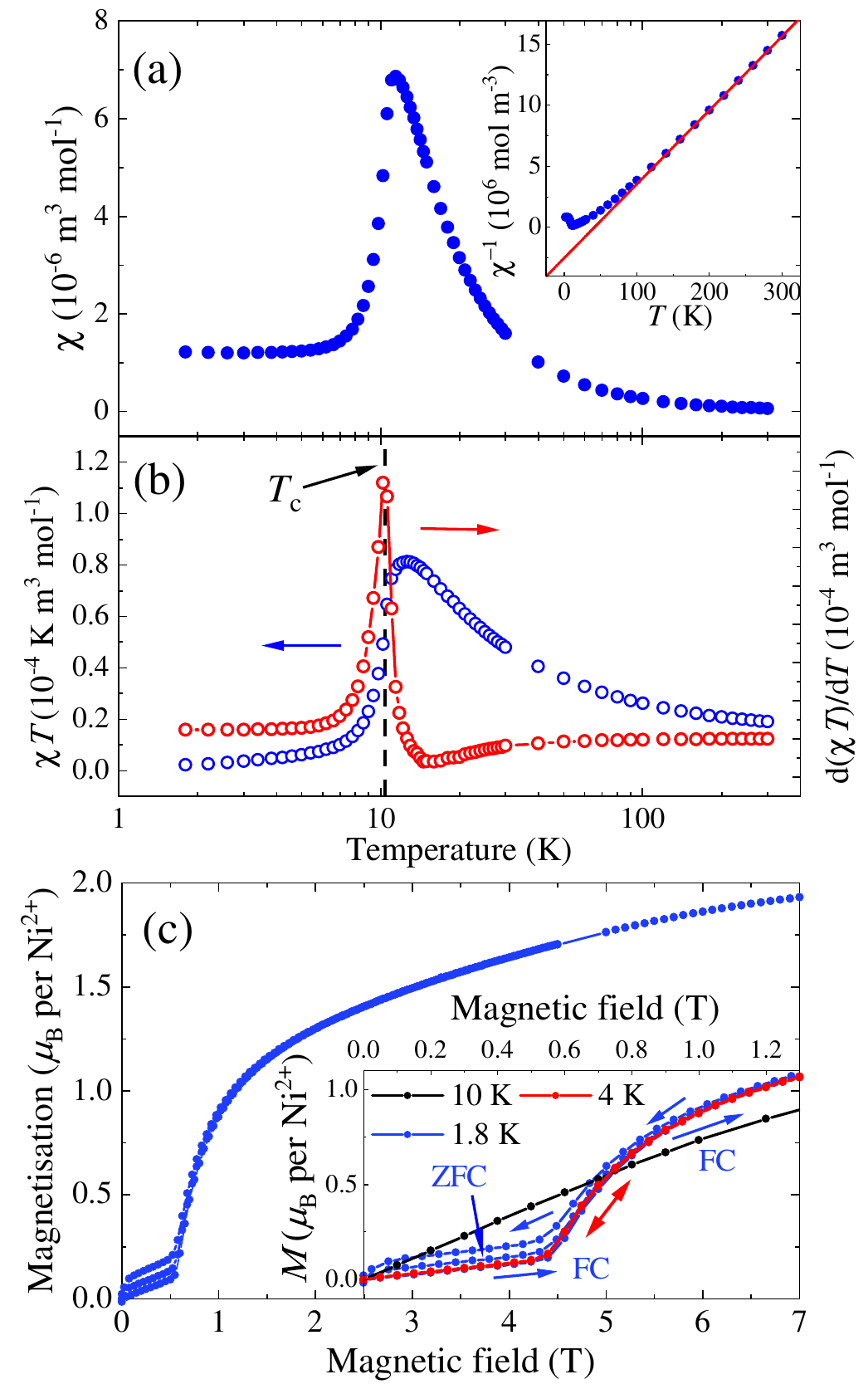}
\caption{\small (a)  Powder magnetic susceptibility $\chi(T)$ data for Ni(NCS)$_{2}$(thiourea)$_{2}$ measured at $\mu_{0}H$ = 0.1 T. Inset shows $\chi^{-1}(T)$ fit to the Curie-Weiss model over the temperature range 100 $\leq T \leq$ 300 K. (b) $\chi(T)$ multiplied by temperature [$\chi T(T)$] (blue, left-axis) and it's derivative (red, right-axis) with critical temperature $T_{\rm{c}}$ marked with a dashed line. (c) Powder DC-field magnetisation $M(H)$ for Ni(NCS)$_{2}$(thiourea)$_{2}$. Inset shows hysteretic behaviour $T < 4~\rm{K}$ with zero-field cooled (ZFC) and field-cooled (FC) sweeps marked.}  \label{fig:Ni_Mag}
\vspace{-0cm}
\end{figure}



Figure \ref{fig:Ni_Mag}(a) shows powder $\chi(T)$ for Ni(NCS)$_{2}$(thiourea)$_{2}$. Upon decreasing temperature, $\chi(T)$ rises to a peak before dropping and plateauing at low-temperatures. The inset of Figure~\ref{fig:Ni_Mag}(a) shows $\chi^{-1}(T)$, where a Curie-Weiss fit for $T \geq100\,\rm{K}$ returns $g = 2.29(1)\,\rm{K}$ and $\Theta_{\rm{CW}} = +42(1)~\rm{K}$. $\chi T(T)$ data, Figure \ref{fig:Ni_Mag}(b), increase on cooling from room temperature, reach a broad maximum and then drop towards zero at $T < 10~\rm{K}$. At high temperature, $\chi T(T)$ data approaches a flat paramagnetic value at $T \approx 300~\rm{K}$, likely plateauing at $T \approx 300-400~\rm{K}$. This corresponds to the energy scale of the largest term in the Hamiltonian $J$, such that $k_{\rm{B}}T \approx 2nJ$ ($n = 2$ is the number of nearest neighbours in the chain). This estimates $J \approx 75-100$\,K, in good agreement with DFT calculations below. The Fisher method \cite{Fisher1962} determines $T_{c} = 10.4(4)~\rm{K}$, Figure \ref{fig:Ni_Mag}(b), in excellent agreement with heat capacity and $\mu^{+}\rm{SR}$ measurements. These observations are consistent with large FM primary exchange interactions and a considerably smaller secondary AFM exchange, leading to an AFM ground state.



Figure \ref{fig:Ni_Mag}(c) shows $M$($H$) for a powder sample of Ni(NCS)$_2$(thiourea)$_2$. At low-fields, $M$($H$) is approximately zero prior to exhibiting a sharp upturn at $\mu_{\rm{0}}H_{\rm sf} = 0.65(5)~\rm{T}$ which we ascribe to spin-flip behaviour. The rise of $M$($H$) slows at fields $\sim 1~$T before increasing monotonically above 3 T, approaching $M$($H$) $\sim$ 2 $\mu_{\rm{B}}$ per Ni$^{2+}$ ion at the maximum experimentally accessible field of 7 T. The field at which the spin-flip occurs can be used to estimate the interchain magnetic exchange $J'$ via $g \mu_{\rm{B}} \mu_{\rm{0}}H_{\rm sf} = 2 S n' J'$ where $n'$ is the number of nearest interchain neighbours \cite{Blackmore2019a}. Taking $g = 2.29(1)$, from $\chi^{-1}(T)$, and $n' = 4$ we estimate $J' = 0.13(1)$ K.

The behaviour of $M$($H$) above $\mu_{\rm{0}}H_{\rm sf}$ can be explained by considering the polycrystalline nature of the sample. Grains with their easy-axis parallel to the applied field are those that contribute to the spin-flip. At $\mu_{\rm{0}}H_{\rm{sf}}$, their spins are rapidly polarised along the field direction. In contrast, grains not orientated with their easy-axis parallel to the field have their spins more slowly polarised along the field direction as the applied field increases. These spins contribute to the slow rise of $M$($H$) after $\mu_{\rm{0}}H_{\rm sf}$. By 7 T, $M$($H$) approaches $\approx$ 2 $\mu_{\rm{B}}$ per Ni$^{2+}$ ion. This is consistent with $M_{\rm{sat}} \approx$ 2.29 $\mu_{\rm{B}}$ per Ni$^{2+}$ ion as expected from the $g$-factor extracted from the fit to $\chi^{-1}(T)$. The inset of Figure \ref{fig:Ni_Mag}(c) shows hysteretic behaviour in $M$($H$) for $T <4$~K similar to that observed in Co(NCS)$_2$(thiourea)$_2$.



\subsection{Calculated exchange coupling constants}
\label{sec:conijs}

In order to help validate the sign and strength of $J$ and $J'$ determined from the magnetometry, the geometries optimised by DFT were used to calculate the magnetic superexchange coupling constants. Along each axis, the energy difference ($\Delta E$) between ferromagnetic (FM) and antiferromagnetic (AFM) coupling was calculated for adjacent $M$ ions and used to obtain the sign and magnitude of the magnetic exchange interaction. As $\Delta E = E_{\rm{FM}} - E_{\rm{AFM}}$, $\Delta E < 0$ is representative of AFM exchange. Values of $\Delta E$ were converted to exchange coupling by considering a single $J$ convention in the Hamiltonian (sum over unique exchange pathways), the results of these calculations are shown in Table \ref{tab:Jcomp} and Table SII \cite{SI}.
We find that the primary exchange is FM in both materials, Table \ref{tab:Jcomp}, with $J = 4.22~\rm{K}$ for the Co species and $J = 78.13~\rm{K}$ for the Ni compound (occurring along $a$, Table SII \cite{SI}). Interchain exchange interactions in both compounds are on the order of $|J'| \sim 0.1~\rm{K}$ and shown in Table \ref{tab:Jcomp}. Weak AFM interchain exchange along $b$ is predicted in both compounds with FM exchange along $c$ predicted in the Ni species as outlined in Table SII \cite{SI}. Such a discrete change in sign of the exchange is difficult to verify experimentally. Due to the small magnitude of $J'$, and the convergence criterion used for the energy calculations, we note that the calculated $J'$ parameters are less reliable than those calculated for $J$. In addition, small changes in the lattice geometry can have statistically significant effects on such small energy differences.
The calculated $J$ is roughly two and four orders of magnitude greater than $J'$ for Co(NCS)$_2$(thiourea)$_2$\ and Ni(NCS)$_2$(thiourea)$_2$ respectively (see Table \ref{tab:Jcomp}); supporting our argument of Q1D behaviour.

\subsection{Charge and spin density}

DFT calculations and magnetometry data both suggest the intrachain exchange in the Co species is significantly weaker than in the Ni species. To investigate the underlying mechanism responsible for the large difference in the values of $J$, electronic configurations of the $M$ sites, charge density maps, and calculated spin density distributions were estimated; see \cite{SI} and \cite{Macchi2011,CrysAlisPRO,Sheldrick,Valkov2016} for details.

High-resolution X-ray diffraction data refined using the Hansen-Coppens multipolar model (MM) \cite{Hansen1978,Holladay1983} was adopted to retrieve the electronic configuration and experimental charge density distribution. The population of the fitted multipolar functions is effectively used to estimate the occupancy of $d$-isorbitals functions. Results for the Co material, Table \ref{tab:nicodpop}, indicate the ion resides in the high-spin [$t_{2g}^5$, $e_g^2$] electronic configuration. The experimental electronic configuration for Co$^{2+}$ in Co(NCS)$_2$(thiourea)$_2$ was validated by comparison with the MM refinement of the structure factor calculated from DFT \cite{Dovesi2014,Peintinger2013}. In addition, DFT simulations allowed estimation of the electronic configuration of Ni$^{2+}$ in Ni(NCS)$_2$(thiourea)$_2$ for which high-resolution X-ray data was not available (Table \ref{tab:nicodpop}). The occupancy of $d$-orbitals functions suggest that the Ni ion is in the [$t_{2g}^6$, $e_g^2$] electronic configuration. In this case, the abundant occupancy of the $d_{x^2-y^2}$ and $d_{z^2}$ orbital functions can be an effect of electron-spin density being partially delocalised onto the ligands. However, an occupancy slightly exceeding the formal one or two electrons is not unusual in MM refinement, since the multipoles are $d$-orbital shape functions freely refined against the structure factor, where occupancy values tending to two electrons represent fully occupied orbitals. The same issue concerns theoretical calculations that use orbital functions to compute a wavefunction from which individual orbital populations are extracted by projecting the crystal orbitals onto atomic basis, and therefore do not guarantee integer occupancy. Here, for sake of a fair comparison, we adopted the very same kind of projection for experiment and calculations, namely, multipolar expansion refined against measured or computed structure factors.

\begin{table}[t]
\small

  \caption{\ Experimentally (Exp. MM) and calculated (Calc. MM) $d$-orbitals occupancy for $M$(NCS)$_2$(thiourea)$_2$ as estimated from the multipolar model (MM). The $z$-axis was set along the unique $M$---N bond whilst the $x$- and $y$-axis were set along $M$---S bonds in the equatorial plane.}
  \label{tab:Jcomp}
  \begin{tabular*}{0.48\textwidth}{@{\extracolsep{\fill}}cccc}
    \hline
& \multicolumn{2}{c}{$M$ = Co} & $M$ = Ni \\\hline\hline
$d$-orbital   & Exp. MM & Calc. MM & Calc. MM \\
\hline
$z^2$         & 1.06(2) & 1.29     & 1.46     \\
$xz$          & 1.34(2) & 1.59     & 2.04     \\
$yz$          & 1.87(2) & 1.64     & 2.06     \\
$x^2-y^2$     & 1.42(2) & 1.30     & 1.44     \\
$xy$          & 1.87(2) & 1.63     & 2.04     \\
tot           & 7.45(10) & 7.47    & 9.05     \\\hline
  \end{tabular*}\label{tab:nicodpop}
\end{table}

\begin{figure}[b]
\centering
	\includegraphics[width=\linewidth]{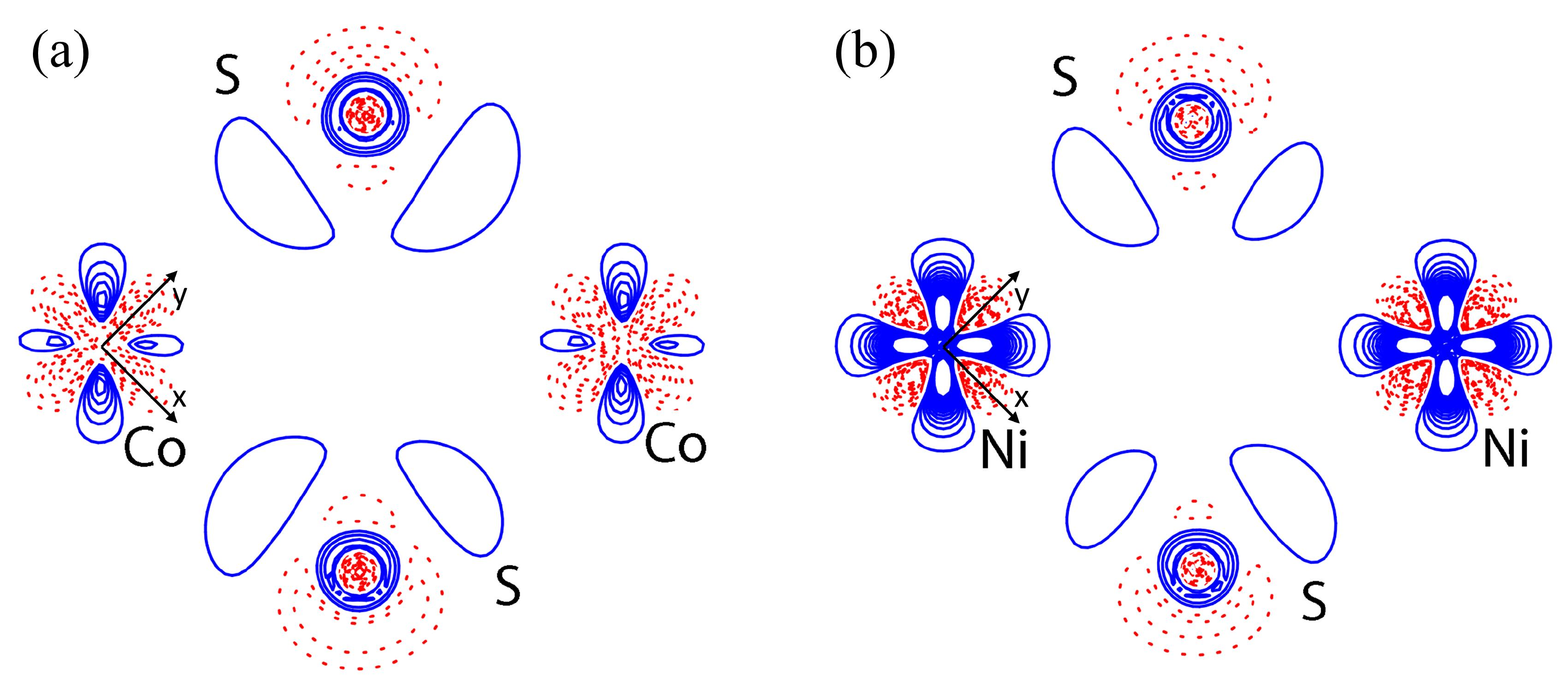}
	\caption{Deformation electron-density maps for (a) Co(NCS)$_2$(thiourea)$_2$ and (b) Ni(NCS)$_2$(thiourea)$_2$. Blue and red contours (0.2 e~\AA$^{-3}$) represent regions of electron charge density excess and depletion, respectively.}
	\label{fig:nicodefden}
\end{figure}

The valence electron-charge density, responsible for chemical bonding, is shown by deformation density maps in terms of regions of charge density excess and depletion obtained from the difference between multipolar and spherical charge density. The different electronic configurations retrieved for Co(NCS)$_{2}$(thiourea)$_{2}$ and Ni(NCS)$_{2}$(thiourea)$_{2}$ are reflected in their deformation density maps. Figure \ref{fig:nicodefden} shows the deformation density in the equatorial plane (relevant for the intrachain magnetic exchange coupling). In both coordination polymers, the electron density is clearly depleted (dotted red contours) towards the S-ligands, the $d_{x^2-y^2}$ being singularly occupied in both materials. Conversely, there is a clear excess of electron-charge density located around the Ni ion in regions between S-ligands (blue contours) depicting the fully occupied $d_{xy}$-orbital. The Co ion has a third semi-occupied orbital pointing between the ligands which leads to a reduction in the electron density along this direction, shown in Figure \ref{fig:nicodefden}(a) as a diminished blue contoured region along the Co $\cdots$ Co (through-space) direction.

Comparing bond lengths alone is not sufficient to establish the equivalence of interactions in isostructural compounds, as different ions have, for instance, different ionic radii which play an important role in defining the strength and nature of the chemical bonds. Therefore, quantum theory of atoms in molecules (QTAIM) \cite{Bader1991} was applied to find bond trajectories and analyse the properties of the electron-charge density at the bond critical points (bcp). Thus, electron-density-based bond properties allow us to quantitatively compare chemical bonds. The electron density and its Laplacian at the bcp have equivalent values in corresponding $M$-ligand interactions Table SV \cite{SI}, establishing that these interactions are indeed analogous in the two materials. Moreover, these quantities emphasize differences between bonds, $e.g.$, the electron density at the $M$---N bond critical point is twice that of the $M$---S bonds confirming strongly anisotropic octahedral environments in both materials. Properties at the $M\cdots M$ bcp are appreciably different, and no bond trajectory and corresponding bcp are found in Ni(NCS)$_2$(thiourea)$_2$. 
Some covalent character of the $M$-ligand interactions is shown by the QTAIM integrated charges in Table SIII \cite{SI}, where the formal oxidation states of the $M^{2+}$ and NCS$^-$ ions are reduced due to charge sharing in the bonds, a clear sign of deviation from purely ionic interactions. Likewise, a slightly positive charge on the thiourea ligand (formally neutral) is indicative of a ligand to metal $\sigma$-donation mechanism.

\begin{figure}[t]
\centering
	\includegraphics[width=\linewidth]{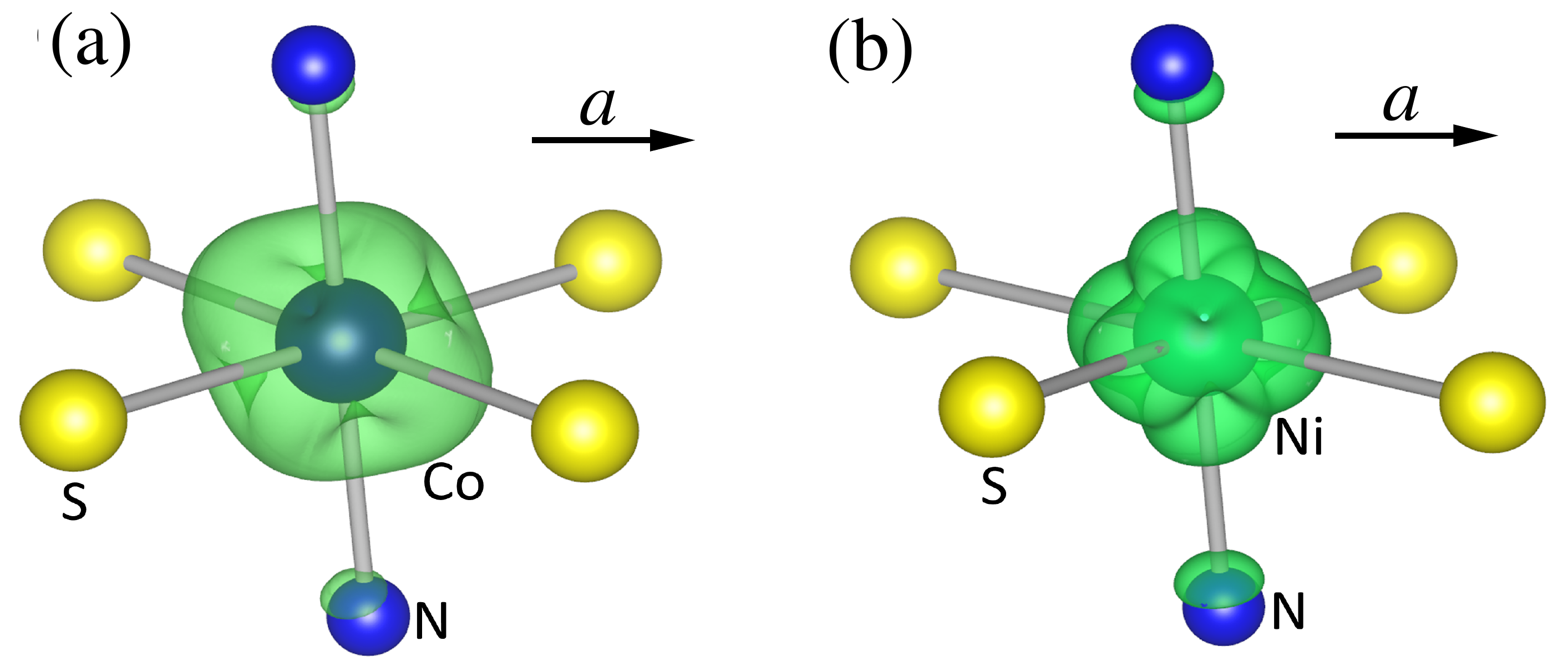}
	\caption{Calculated spin-density maps for (a) Co(NCS)$_2$(thiourea)$_2$\ and (b) Ni(NCS)$_2$(thiourea)$_2$\ plotted at the isovalue of 0.005\,$e$ per \AA$^{3}$. Spin-density regions are outlined as 3D green contours.}
	\label{fig:nicosd}
\end{figure} 

Electronic configurations and deformation density maps inferred the possible magnetic superexchange pathways in the materials, establishing that in Co(NCS)$_{2}$(thiourea)$_{2}$, a through-space magnetic interaction is available, whereas in Ni(NCS)$_{2}$(thiourea)$_{2}$ it is absent. The calculated electron spin-density distributions for the two compounds follow explicitly the differences in their electronic configurations. In Co(NCS)$_2$(thiourea)$_2$, the calculated spin-density distribution, centred around the Co ion, is quite spherical as shown in Figure \ref{fig:nicosd}(a). Here, the spin-density is concentrated not only towards the ligands but also directly along the $a$-axis, permitting two possible exchange pathways between adjacent Co ions: a through-bond interaction along the two Co---S---Co bridges and a through-space Co $\cdots$ Co exchange interaction. The latter must be AFM as it is a result of bond-formation which is subject to the Pauli exclusion principle. In contrast, the spin-density at the Ni sites is polarised only along the ligands, see Figure \ref{fig:nicosd}(b). In this case, the spin-density is concentrated exclusively along the Ni---S---Ni pathways and towards the Ni---N bond. Therefore, Ni(NCS)$_2$(thiourea)$_2$ can only exploit the Ni---S---Ni through-bond interactions and, unlike Co(NCS)$_2$(thiourea)$_2$, has no through-space magnetic exchange.

\section{Discussion}

\begin{table}[b]
\small

  \caption{\ Comparison of the intrachain $J$, interchain $J'$, and SIA $D$ determined experimentally [$a=\chi(T)$, $b = M(H)$, $c = C_{\rm mag}(T)$] and via DFT calculations. Negative exchange values indicate AFM interactions and negative $D$ values indicate easy-axis behaviour. DFT values were calculated by considering a single \textit{J} convention in the Hamiltonian where summations are made over unique exchange pathways.}
  \label{tab:Jcomp}
  \begin{tabular*}{0.48\textwidth}{@{\extracolsep{\fill}}ccccccc}
    \hline
     &\multicolumn{3}{c}{Co(NCS)$_2$(thiourea)$_2$}&
     \multicolumn{3}{c}{Ni(NCS)$_2$(thiourea)$_2$} \\ 
    \hline

& $J$ (K) & $J'$ (K) & $D$ (K)  & $J$ (K) & $J'$ (K) & $D$ (K) \\ \hline
Exp.  & 3.62(7)$^a$ &   $-0.3(1)^b$ & $\sim -100$$^a$ & $\sim 100$$^a$ & $-0.13(1)^b$ & $\sim-10^c$ \\
DFT  & 4.22 &  $-0.1$ &  --- &   78.13 &   $-0.1$  & --- \\
\hline
  \end{tabular*}
\end{table}

Table \ref{tab:Jcomp} shows the values of $J$, $J'$ and $D$ determined experimentally for both the Ni and Co compounds. An axial compression of the $M$S$_4$N$_2$ octahedra results in Ising-like anisotropy in both materials. As is typical, see \cite{Boca2004}, the anisotropy energy in the Co material is found to be considerably larger than in the Ni system.

Both compounds exhibit FM intrachain exchange, which is an order of magnitude greater in the Ni material than in the Co system. The electronic configurations, by identifying the semi-occupied (magnetic) orbitals shown in Table \ref{tab:nicodpop}, determine that in Ni(NCS)$_2$(thiourea)$_2$ only a through-bond $M$---S---$M$ exchange is possible, whereas in Co(NCS)$_2$(thiourea)$_2$ both through-bond $M$---S---$M$ and through-space $M\cdots M$ magnetic interactions are available. As a result, in both materials spin-density extends along the thiourea bibridges (Figure \ref{fig:nicosd}), giving rise to strong FM exchange along the $M$---S---$M$ pathways. However, the spin-density in the Co compound also extends directly along the $a$-axis, which leads to an additional AFM through-space interaction between Co ions. Competition between the two exchange pathways reduces the resultant intrachain exchange and explains the lower FM $J$ in the Co system. This highlights the sometimes subtle ways in which the choice of transition metal ion affects the effective exchange strength. DFT calculations estimate the average interchain exchange to be around $-0.1~\rm{K}$ in both compounds in good agreement with magnetometry measurements.

\sloppypar
Ni(NCS)$_2$(thiourea)$_2$ can be compared to the related material NiCl$_2$(thiourea)$_4$ (DTN), an AFM-coupled spin chain that contains two field-induced phase transitions that may belong to the universality class of BEC~\cite{Zapf2006a,Figgis1986}. Exchanging the Cl$^-$ ion for the NCS$^-$ ion causes a structural change from $I$4 for DTN~\cite{Figgis1986} to $P\overline{1}$ in the present case, and the change in the local Ni(II) environment, from NiS$_4$Cl$_2$ to NiS$_4$N$_2$, has a strong effect on the SIA. The equatorial Ni---S bond lengths are similar in both systems: 2.46 \AA\ for DTN and an average value of 2.57(3)~\AA\ for Ni(NCS)$_2$(thiourea)$_2$. However, the octahedra in DTN have a slight axial elongation with an equatorial to axial bond-length ratio of 0.98, while in Ni(NCS)$_2$(thiourea)$_2$ the octahedra are axially compressed with a bond-length ratio of 1.26. This results in the Ising-like SIA of $D \sim -10~\rm{K}$, compared to the easy-plane SIA of +8.12(4)~K observed in DTN \cite{Zapf2006a}. 

The lack of spatial-inversion-symmetry in DTN results in a net electric polarisation along the $c$ direction \cite{Mun2014}. In contrast, Ni(NCS)$_2$(thiourea)$_2$ does possess a centre of inversion symmetry, so we expect no net electric polarisation.

The structure and SIA of DTN is such that AFM exchange is mediated along linear Ni---Cl$~\cdots~$Cl---Ni pathways [$J$ =  -1.74(3) K] propagating along the $c$-axis. Within the $ab$-plane, non-bridging thiourea ligands keep the magnetic ions well separated (adjacent Ni sites are 9.595 \AA~apart), which results in weak AFM $J' = -0.17(1)~\rm{K}$, possibly taking advantage of Ni---Cl$\cdots$(H$_2$N)$_2$---C---S---N superexchange pathways. 

For Ni(NCS)$_2$(thiourea)$_2$, the thiourea ligands are no longer terminal but now connect adjacent Ni sites along $a$ via Ni---S---Ni bibridges which mediate the large FM exchange along $a$.
The thiourea ligands still keep adjacent Ni sites well separated along $b$ at 7.527\,\AA, and H-bonding between the ligands mediates weak AFM $J'$. Although Ni---NCS---Ni bond pathways have been shown to effectively mediate magnetic exchange interactions in similar compounds \cite{Palion-Gazda2015}, we find here that the NCS ligands are terminal and support only weak AFM $J'$ via H-bonds along $c$. Thus, whilst DTN and Ni(NCS)$_2$(thiourea)$_2$ have highly similar equatorial environments, the dramatic structural change, invoked by substitution of the axial Cl$^-$ for NCS$^-$, leads to drastically different magnetic properties; from a Q1D $XY$-like AFM ground-state exhibiting QCP behaviour (DTN), to that of an FM coupled chain of Ising spins in Ni(NCS)$_2$(thiourea)$_2$.

Similarly, Co(NCS)$_2$(thiourea)$_2$ can be compared to the $S$ = 3/2 analogue of DTN, CoCl$_2$(thiourea)$_4$ (DTC) \cite{DTCstruc}. DTC
displays antiferromagnetic order below approximately 1~K that can be suppressed by magnetic fields of around 2~T, and somewhat surprisingly, the observed susceptibility and magnetisation are largely isotropic~\cite{Mun2014,Forstat1966}.
Swapping the Cl$^-$ for NCS$^-$ ion results in a structural change, now from $P 4_2 / n$ to $P \overline{1}$. Average Co---S bond lengths in DTC and Co(NCS)$_2$(thiourea)$_2$ are equal to within errors at 2.53(3)\,\AA and 2.57(3)\,\AA, respectively. The local CoS$_4$N$_2$ environment is compressed along the axial N---Co---N bond such that the ratio of the equatorial Co---S to axial Co---N bond is 1.28, leading to a large Ising-like SIA $D \sim -100~\rm{K}$. This contrasts with the isotropic behaviour seen in DTC which also possesses a slight axial compression, with a bond-length bond ratio of 1.02. Analogous to the comparison of the Ni species and DTN above, thiourea ligands form Co---S---Co bibridges in Co(NCS)$_2$(thiourea)$_2$ that mediate strong FM exchange along the $a$-axis, while the chains are well separated along $b$. Again NCS ligands are terminal and mediate weak AFM interactions along via H-bond networks along the $c$-axis.

Co(NCS)$_2$(thiourea)$_2$ can also be compared to the archetype transverse-field Ising chain material CoNb$_2$O$_6$, with both possessing strong Ising-like SIA with FM $J$ and weak AFM $J'$ \cite{Kinross2014}. Interchain exchange interactions induce a transition to long-range AFM order in both, at $T_{\rm{c}} = 2.9~\rm{K}$ in CoNb$_2$O$_6$ \cite{Scharf1979} and $T_{\rm{c}} = 6.82(5)~\rm{K}$ in Co(NCS)$_2$(thiourea)$_2$.
The lower $T_{\rm{c}}$ for CoNb$_2$O$_6$ indicates a more ideal 1D system, with $J' \sim 0.01~\rm{K}$ \cite{Kunimoto1999} compared to 0.1~K for Co(NCS)$_2$(thiourea)$_2$.
In CoNb$_2$O$_6$, a critical field applied perpendicular to the Ising-axis breaks the 3D-AFM order, pushing it through a QCP as the material enters a quantum paramagnetic state~\cite{Kinross2014,Coldea2010}. Because application of a transverse field consistently shattered Co(NCS)$_2$(thiourea)$_2$ single-crystals, we are as yet unable to provide evidence of similar quantum-critical behaviour in our material.

\section{Conclusion}

In summary, we find that $M$(NCS)$_2$(thiourea)$_2$, where $M = $ Ni(II) or Co(II), both behave as Q1D chains with ferromagnetic intrachain exchange $J$, weak antiferromagnetic interchain interactions $J'$ and Ising-like single-ion anisotropy ($D < 0$). At low temperature, long-range AFM ordering is observed in both materials as confirmed by heat capacity, magnetometry and $\mu^+$SR measurements. The considerable difference in the magnitude of $J$ between the two compounds is due to their electronic configurations, where semi-occupied orbitals are responsible for the different spin-density distributions, highlighting the prominent role of the transition metal-ion in promoting Q1D behaviour. We find that the magnetic properties of the materials discussed here are very different to those observed in the chemically related quantum magnets DTN and DTC, owing to significant structural changes induced by substitution of the axial Cl$^-$ halide ion for the NCS$^-$ ion.

\section*{Acknowledgments}
This project has received funding from the European Research Council (ERC) under the European Union’s Horizon 2020 research and innovation program (Grant Agreement No. 681260). Work at EWU was supported by NSF Grant No. DMR-1703003. NSF’s ChemMatCARS Sector 15 is principally supported by the Divisions of Chemistry (CHE) and Materials Research (DMR), NSF, under Grant No. NSF/CHE-1834750. R. S. acknowledges support from the Swiss National Science Foundation (project Nr.P2 BEP2$\_$188253).This work is supported by the EPSRC (under grants EP/N023803/1 and EP/N024028/1). The National  High Magnetic Field Lab is funded by the U.S. National Science Foundation,  US Department of Energy and the State of Florida through  Cooperative  Grant  No.  DMR-1157490. Part of this work was carried out at the Swiss Muon Source, Paul Scherrer institut, Switzerland and we are grateful to Hubertus Luetkens for experimental assistance. F.X. would like to acknowledge the funding from the European Union's Horizon 2020 research and innovation program under the Marie Sk\l{}odowska-Curie grant agreement No 701647. We thank Janice Musfeldt for useful discussions. Data presented in this paper resulting from the UK effort will be made available at https://wrap.warwick.ac.uk/148301.


\providecommand{\noopsort}[1]{}\providecommand{\singleletter}[1]{#1}%

\end{document}


\title{Magnetic ground-state of the highly one-dimensional ferromagnetic \\chain compounds \textit{M}(NCS)$_2$(thiourea)$_2$; \\ \textit{M} = Ni, Co - Supplemental information.}

\author{S. P. M. Curley}
\affiliation{Department of Physics, University of Warwick, Gibbert Hill Road, Coventry, CV4 7AL, UK}
\author{R. Scatena}
\affiliation{Department of Physics, Clarendon Laboratory, University of Oxford, Parks Rd., Oxford, OX1~3PU, United Kingdom}
\affiliation{Department of Chemistry and Biochemistry, University of Bern, 3012 Bern, Switzerland}
\author{R. C. Williams}
\author{P. A. Goddard}
    \email{p.goddard@warwick.ac.uk}
\affiliation{Department of Physics, University of Warwick, Gibbert Hill Road, Coventry, CV4 7AL, UK}
\author{P. Macchi}
\affiliation{Department of Chemistry, Materials, and Chemical Engineering, Polytechnic of Milan, Milan 20131, Italy}
\affiliation{Istituto Italiano di Tecnologia, Center for Nano Science and Technology CNST@polimi, Milan 20133, Italy}
\affiliation{Department of Chemistry and Biochemistry, University of Bern, 3012 Bern, Switzerland}
\author{T. J. Hicken}
\author{T. Lancaster}
\affiliation{Durham University, Department of Physics, South Road, Durham, DH1~3LE, United Kingdom}
\author{F. Xiao}
\affiliation{Laboratory for Neutron Scattering and Imaging, Paul Scherrer Institut, CH-5232 Villigen PSI, Switzerland}
\author{S. J. Blundell}
\affiliation{Department of Physics, Clarendon Laboratory, University of Oxford, Parks Road, Oxford, OX1~3PU, United Kingdom}
\author{V. Zapf}
\affiliation{National High Magnetic Field Laboratory, Materials Physics and Applications Division, Los Alamos National Laboratory, Los Alamos, New Mexico 87545, USA}
\author{J. C. Eckert}
\author{E. H. Krenkel}
\affiliation{Physics Department, Harvey Mudd College, Claremont, CA, United States}
\author{J. A. Villa}
\author{M. L. Rhodehouse}
\author{J. L. Manson}
    \email{jmanson@ewu.edu}
\affiliation{Department of Chemistry and Biochemistry, Eastern Washington University, Cheney, Washington 99004, USA}

\maketitle

\begin{sloppypar}

\renewcommand{\thefigure}{S\arabic{figure}}
\renewcommand{\thetable}{S\Roman{table}}

\noindent
\section{Experimental details}

\noindent
\subsection{Synthesis}

All chemical reagents were obtained from commercial sources and used as received.  All reactions were performed in 50 ml glass beakers.  For Ni(NCS)$_2$(thiourea)$_2$, an aqueous solution of NiCl$_2$ (0.3018g, 1.262 mmol) was slowly added to an aqueous solution of NH$_4$NCS (0.2113g, 2.7764 mmol) and thiourea (0.1918g, 2.524 mmol) to afford a clear green solution.  When allowed to slowly evaporate at room temperature, X-ray quality, dark green, fine, needle-like crystals were recovered.  For Co(NCS)$_2$(thiourea)$_2$, an aqueous solution of CoCl$_2$ (0.29996g, 1.2607 mmol) was slowly added to an aqueous solution of NH$_4$NCS (0.2111g, 2.7737 mmol) and thiourea (0.1919g, 2.5210 mmol) to afford a slightly cloudy, pink-red solution.  When allowed to slowly evaporate at room temperature, X-ray quality, red-brown, large, needle-like crystals were recovered.

\subsection{Muon Spin Relaxation}

In a positive muon spin relaxation ($\mu^+$SR) experiment, positive spin-polarised muons are implanted in a sample where they interact with the local magnetic field.
Static magnetic order will typically cause coherent precession of the muon-spin with angular frequency $\omega_\mu = \gamma_\mu B$, where $\gamma_\mu$ is the muon gyromagnetic ratio and $B$ is the local field at the muon site, which is a sum of the internal field arising due to the spin configuration, and the external field $\mu_{0}H$. Alternatively, if the system is dominated by dynamic magnetic effects, the muon-polarisation will typically decay exponentially with relaxation rate $\lambda = 2\Delta^2\tau$, where $\Delta$ is the width of the field at the muon site and $\tau$ is the correlation time of the local field.



After, on average, 2.2~$\mu$s, the muon decays into a positron and two neutrinos, where the positron is preferentially emitted along the direction of muon-spin polarisation at the time of decay.
By detecting these positrons one can calculate a quantity directly proportional to the muon-spin polarisation, known as the asymmetry,
\noindent
\begin{equation}
	A(t) = \frac{N_\text{F}(t)-\alpha N_\text{B}(t)}{N_\text{F}(t)+\alpha N_\text{B}(t)} ,
\end{equation}
\noindent
where $N_\text{F,B}(t)$ are the number of muons detected in the detectors forward and backward of the initial muon-spin polarisation respectively, and $\alpha$ is an experimental efficiency factor.

Zero field (ZF)-$\mu^+$SR measurements on two polycrystalline compounds, $M$(NCS)$_2$(thiourea)$_2$ where $M$ = Ni or Co, were made using the GPS instrument at Swiss Muon Source (S$\mu$S), Paul Scherrer Institut (PSI).
The samples were packed in a silver foil packet (thickness 25~$\mu$m) and mounted on the end of a fork in a $^4$He Quantum Cryostat.
Data analysis was performed using both WiMDA~\cite{pratt2000wimda} and our own software developed for global fitting.

\subsection{Heat capacity}

Heat capacity measurements were made on a Quantum Design PPMS instrument. Plate-like single-crystals were available for Co(NCS)$_2$(thiourea)$_2$, whilst polycrystalline powder samples of Ni(NCS)$_2$(thiourea)$_2$ were pressed into pellets. The samples were mounted onto the sample stage with a small amount of Apiezon vacuum M-grease to provide good thermal contact between the sample and sample platform. Addenda measurements were performed to correct for the heat capacity of the sample platform and M-grease. The mounted sample initially sits in thermal equilibrium with the platform. A small heat pulse is then applied and the time taken for the system to relax back to thermal equilibrium monitored. The heat capacity can then be extracted by fitting the resultant relaxation curve \cite{Lashley2003}.

\subsection{Magnetometry}

Magnetometry measurements were performed using a Quantum Design MPMS-7XL SQUID magnetometer. Ni(NCS)$_{2}$(thiourea)$_{2}$ was crushed to a fine powder and mixed with Vaseline to mitigate grain reorientation in the applied field. The Vaseline-sample mix was mounted within a low-magnetic-background gelatin capsule. A plate-shaped single-crystal of Co(NCS)$_{2}$(thiourea)$_{2}$ was mounted on a low-magnetic-background Teflon sample holder in order to measure along all multiple axes of the single-crystal. Both were then placed inside a plastic drinking straw to be measured within the MPMS sample chamber. Temperature sweeps from $1.8 \leq T \leq 300~\rm{K}$ were completed on both compounds. Field sweeps up to $\mu_{\rm{0}} H = 7~\rm{T}$ were completed for Ni(NCS)$_{2}$(thiourea)$_{2}$. For Co(NCS)$_{2}$(thiourea)$_{2}$, application of sizable magnetic fields perpendicular to the Co(II) Ising-axis regularly shattered or re-orientated the samples, regardless of the method used to affix them.

\subsection{Charge and spin density}

Single-crystal X-ray measurements were performed on Co(NCS)$_{2}$(thiourea)$_{2}$ and Ni(NCS)$_{2}$(thiourea)$_{2}$ using an Oxford Diffraction SuperNova area-detector diffractometer. An Al-filter was used to reduce the number of low-energy photons contaminating the incident Mo K$_{\alpha}$ radiation ($\lambda=$0.71073\,\AA) \cite{Macchi2011}. Single-crystals were adhered to a glass fibre and low-temperatures achieved using an Oxford Cryostream flow cryostat. Extensive high-resolution data collections ($d_{\rm{min}} = 0.50$~\AA) were performed at 100.0(2) K for Co(NCS)$_{2}$ (thiourea)$_{2}$; Ni(NCS)$_{2}$ (thiourea)$_{2}$ data was collected at 173.0(2) K, up to a resolution of $0.70$~\AA. Due to Ni(NCS)$_{2}$(thiourea)$_{2}$ single-crystals being irremediably affected by twinning, high quality X-ray data collection, required for charge density modelling, could only be performed for Ni(NCS)$_{2}$(thiourea)$_{2}$. Data reductions were performed using CrysAlisPro software \cite{CrysAlisPRO}. Intensities were analytically corrected for Lorentz, polarisation and absorption effects using the Gaussian method, based on the dimensions of the Co(NCS)$_{2}$(thiourea)$_{2}$ single-crystals. For Ni(NCS)$_{2}$(thiourea)$_{2}$, a semi-empirical correction was applied. The conventional structural refinements were carried out with SHELXL \cite{Sheldrick}.

For charge density analysis of Co(NCS)$_2$(thiourea)$_2$, the atomic and thermal parameters obtained from spherical refinement were used as a starting point for the Multipolar Model refinement within the Hansen-Coppens formalism \cite{Hansen1978} for electron charge density determination. Calculations were performed using XD2016 \cite{Valkov2016}. All multipoles of non-hydrogens atoms were refined up to the hexadodecapole. $\kappa$ and $\kappa^\prime$ parameters were also freely refined except for the S atom where they were constrained to vary equally. Hydrogen atoms were restrained to calculated distances and refined using a riding model. Only the monopole and a dipole oriented in the bond direction were refined for the hydrogens. Both $\kappa$ and $\kappa^\prime$ were constrained to 1.3 and 1.2, respectively. Residual density maps show small and rather random discrepancies. Topological properties of the electron density, integrated atomic charges and electrostatic potential were calculated with the XDPROP module of XD software.

Periodic Density Functional Theory (p-DFT) calculations of geometry optimizations and electron density distributions for both compounds were carried out using CRYSTAL14 software \cite{Dovesi2014}. The functional B3LYP was used with basis pob-TZVP \cite{Peintinger2013}. The theoretical electron density was analyzed with the TOPOND module of CRYSTAL14. Computed structure factors were refined with the multipolar model using XD2016 \cite{Valkov2016}.

\section{Results}

\subsection{Muon Spin Relaxation}

Spectra measured in the magnetically ordered regime for Ni(NCS)$_2$(thiourea)$_2$, were fitted with
\noindent
\begin{multline}
	A(t) = \sum_{i=1}^2 A_i e^{-\lambda_i t}\cos(2\pi\nu_i t + \phi_i)\\
	+ A_3 e^{-\lambda_3 t} + A_\text{bsln} e^{-\lambda_\text{bsln} t} ,
\end{multline}
\noindent
where $A_i$ are the relaxing asymmetry of each component with corresponding relaxation rate $\lambda_i$, oscillation frequency $\nu_i$, and the phase $\phi_i$.
The amplitude $A_\text{bsln}$ is the baseline asymmetry (accounting for both muons stopping outside of the sample and for those that stop within the sample but only have a component of their spin along the local field direction) which has relaxation rate $\lambda_\text{bsln}$.
Each component of the model accounts for muons that stop in different
local environments, with the two oscillatory components accounting for muons that stop in two different local magnetic fields that lead to coherent precession, indicative of long range magnetic order.

To account for the spectra measured at $T~\lesssim~7$~K, which showed a third frequency at approximately $3\nu_1$, a purely relaxing component was included in the fitting to account for the rapid depolarisation of the initial muon-spin due to this third precession rate.
As expected, the relaxation rate of this component $\lambda_3$ is large at low temperatures and drops for $T~\gtrsim~7$~K.
This indicates a broad transition region, similar to the observation made on Co(NCS)$_2$(thiourea)$_2$ in the main text.

Above $T_\text{N}$, where oscillations are not resolvable in the
spectra, the data are well described by a model comprising of two relaxing components,
\noindent
\begin{equation}\label{eqn:tworel}
	A(t) = \sum_{i=1}^2 A_i e^{-\lambda_i t} + A_\text{bsln} e^{-\lambda_\text{bsln} t} .
\end{equation}
\noindent
The background relaxation rate $\lambda_\text{bsln}$ was set to be equal to that found from fitting below $T_\text{N}$, and all other parameters were globally refined to be constant over the entire temperature range, except for the small relaxation rate $\lambda_1$, which was found to decrease on increasing temperature.

For the ZF-$\mu^+$SR asymmetry spectra measured for Co(NCS)$_2$(thiourea)$_2$, Eqn.~\ref{eqn:tworel} was fitted over the entire temperature range.
The background relaxation rate $\lambda_\text{bsln}$ and the large relaxation rate $\lambda_2$ were globally refined over all the datasets to obtain the best fits.

\subsection{Heat Capacity}

\begin{table}[b]
\centering
\caption{Fitted high-temperature contribution to the heat capacity for  Co(NCS)$_2$(thiourea)$_2$ and Ni(NCS)$_2$(thiourea)$_2$. $A_{i}$ is amplitude of mode $i$, whilst $\theta_{i}$ is the characteristic temperature of the mode $i$ ($i$ = D, Debye; $i$ = E, Einstein).}{
\label{tab:Latt_fit}
\begin{small}
\begin{tabular}{lrr}
\multicolumn{1}{l}{Parameter}   & Co(NCS)$_2$(thiourea)$_2$      & Ni(NCS)$_2$(thiourea)$_2$     \\\hline\hline
$A_{\rm{D}}$ (J/K.mol)                   & 94(4)       & 172(5)      \\
$\theta_{\rm{D}}$ (K)                    & 172(3)       & 443(28)      \\
$A_{\rm{E_{1}}}$ (J/K.mol)               & 105(3)       & 96(8)      \\
$\theta_{\rm{E_{1}}}$ (K)                & 285(7)       & 126(5)     \\
$A_{\rm{E_{2}}}$ (J/K.mol)               & 172(3)       & 185(6)      \\
$\theta_{\rm{E_{2}}}$ (K)                & 816(13)       & 941(46)      \\
           
\end{tabular}
\end{small}}
\end{table}

\begin{figure}[t]
\centering
\includegraphics[width= 1 \linewidth]{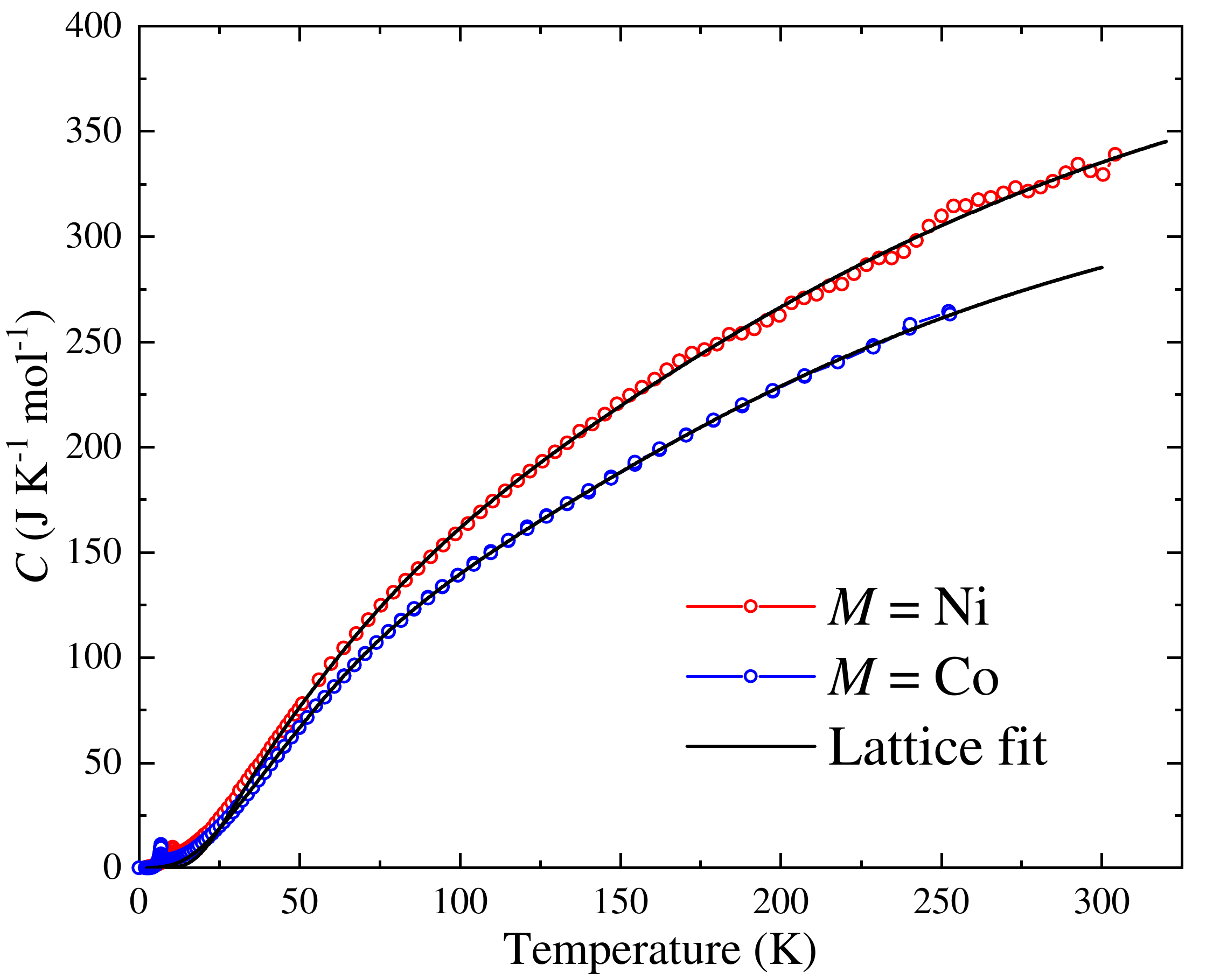}
\caption{\small Heat capacity ($C$) measured in zero-field and plotted as a function of temperature for Co(NCS)$2$(thioures)$2$ and Ni(NCS)$2$(thiourea)$2$, solid lines are fits to two Einstein and one Debye mode for the temperature range $T > 30~\rm{K}$ and $T > 37~\rm{K}$, respectively. Fit parameters are shown in Table \ref{tab:Latt_fit}.}\label{fig:Hc_latt_fit}
\vspace{-0cm}
\end{figure}

Figure \ref{fig:Hc_latt_fit} shows zero-field heat capacity ($C$) divided temperature ($T$) and plotted as a function of $T$ for both compounds. Both $C/T$ datasets show sharp peaks at $T_{c} = 6.82(5)$~K and 10.5(1)~K for the Co(II) and Ni(II) respectively, indicative of long-range ordering. To isolate the low-temperature magnetic contribution to the heat capacity, the high-temperature ($T \gtrsim 30$~K) contribution ($C_{{\rm high}T}$) can be reproduced by a model containing two Einstein and one Debye phonon mode, with the resultant fit subtracted as a background ($C_{\rm mag} = C-C_{{\rm high}T}$) as described in \cite{Blackmore2019a}. The black lines in Figure~\ref{fig:Hc_latt_fit} are fits to $C_{{\rm high}T}$, the resultant fit parameters are shown in Table~\ref{tab:Latt_fit}. The model used to fit the high-temperature heat capacity is purely phenomenological and is not intended to provide an accurate parameterization of the lattice dynamics.

\begin{figure}[t]
\centering
\includegraphics[width= 1 \linewidth]{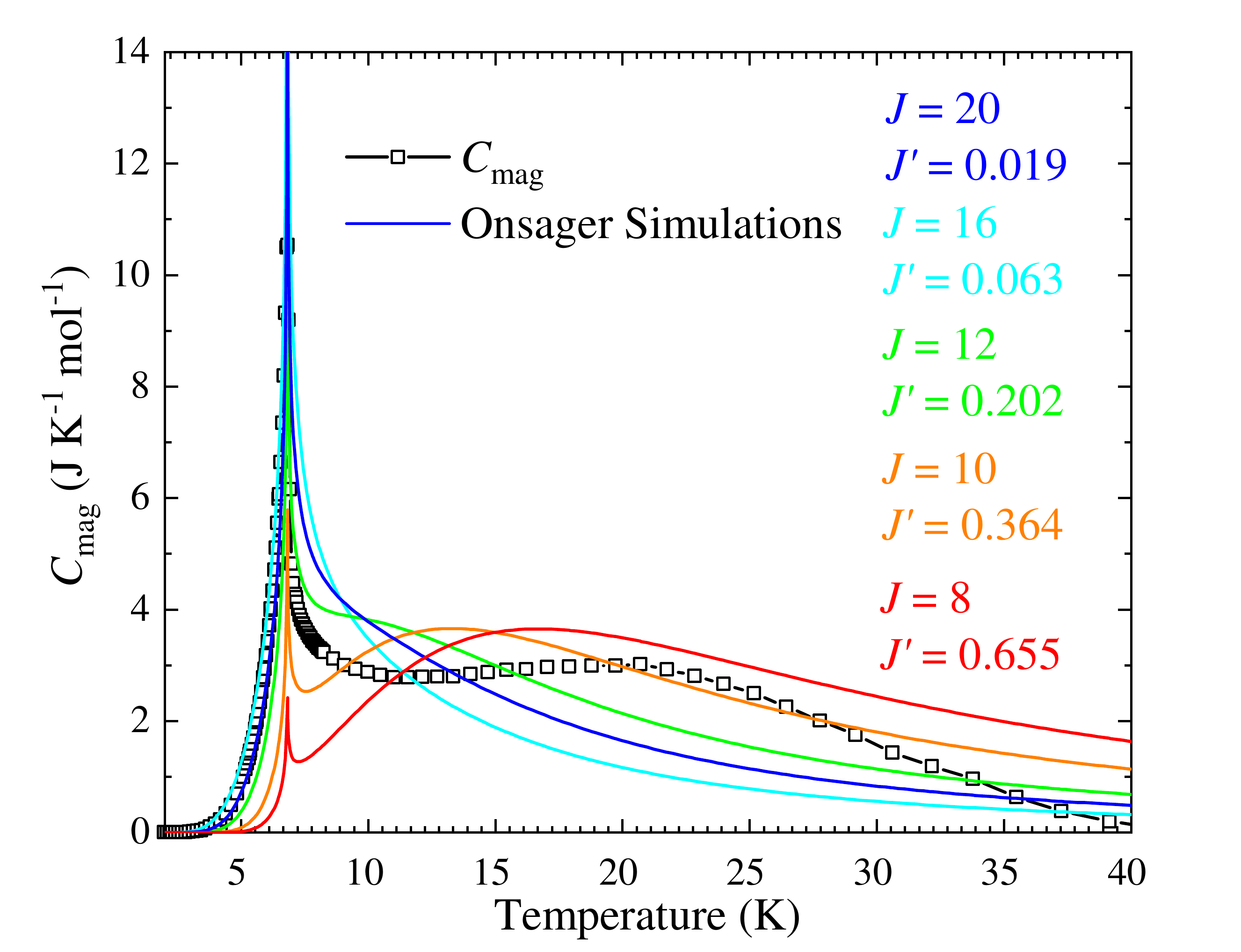}
\caption{\small Magnetic heat-capacity $C_{\rm{mag}}$ (squares) plotted as a function of temperature $T$. Simulated $C_{\rm{mag}} (T)$ curves (solid lines) are plotted for a selection of exchange values $\{J,J'\}$ which obey the Onsager relation (Eq. \ref{EqSinhrule}) for a 2D lattice of Ising spins.}\label{Onsager_plot}
\vspace{-0cm}
\end{figure}

For the Co(II) species, the large negative single-ion anisotropy
(SIA) term splits the two Kramer doublets such that, at low temperatures,
the system can be well approximated by the effective
spin-half Ising model. The sharp nature of the peak in $C_{\rm{mag}}$ rules
out strictly 1D behaviour, where spin-spin correlations begin to
build up well before $T_c$, resulting in a broad transition. For a
2D square lattice of Ising-spins with anisotropic exchange interactions
($J$, $J'$), the $C_{\rm{mag}}$ response can be simulated by considering
Eq. \ref{EqSinhrule}, as derived by Onsager \cite{Onsager1944,McCoy1973}, where $J$ and $J'$ obey:
\noindent
\begin{equation}\label{EqSinhrule}
    \sinh(\frac{2J}{T_c}) \sinh(\frac{2J'}{T_c}) = 1
\end{equation}
\noindent
Here, $J$ and $J’$ are the magnetic exchange interactions along two
different directions within the 2D lattice. Taking $T_c = 6.82(5)~\rm{K}$, $C_{\rm{mag}}(T)$ can be simulated and compared to the measured response
for various sets of $\{ J,J' \}$ which obey Eq. \ref{EqSinhrule}. The simulated
curves (solid lines) were then compared to the measured response
for Co(NCS)2(thiourea)2 (open squares) with the results shown
in Figure \ref{Onsager_plot}.

The simulated curves do not capture the data convincingly,
leading to the conclusion that $C_{\rm{mag}}$ cannot be well described by
the 2D Ising model. We interpret this result as evidence to support
a 3D Ising long range ordering within the Co compound. The two
classes of interchain exchange interaction are therefore expected
to be comparable in magnitude, and the number of interchain
next nearest neighbours expected to be $n' = 4$.

\noindent
\noindent
\subsection{Charge and spin density}

\begin{table}[b]
\centering
\caption{Energy differences ($\Delta E$) and calculated magnetic super-exchange coupling constants ($J$) between nearest neighbour transition metal-ions along the principle crystallographic axes. Conversion from $\Delta E$ to \textit{J} is done by considering a single \textit{J} convention. Negative values describe antiferromagnetic interactions.}{
\label{tab:nicoj}
\begin{small}
\begin{tabular}{ccccc}
              & \multicolumn{2}{c}{Co(NCS)$_2$(thiourea)$_2$}     &  \multicolumn{2}{c}{Ni(NCS)$_2$(thiourea)$_2$}   \\\hline\hline
Axis			   & $\Delta E$ (K) & $J$ (K)  & $\Delta E$ (K) & $J$ (K) \\ \hline
$a$  & 25.31 &   4.22           & 234.38 & 78.13  \\
$b$  & -0.61 &  -0.10           &  -0.34 &  -0.11  \\
$c$  & -0.51 &  -0.08           &   0.20 &   0.07  \\           
\end{tabular}
\end{small}}
\end{table} 
\noindent
Selected structural refinement parameters for single-crystal X-ray refinement of Co(NCS)$_2$(thiourea)$_2$ and Ni(NCS)$_2$(thiourea)$_2$ are shown in Table \ref{tab:xrayCoNi}, where the lack of twinning in the Co species allowed high-quality multipolar model (MM) refinement. The MM refinements were used to calculate the electron deformation density maps (Figure 9, main text) by comparing it to that calculated by a simple Independent Atom Model (IAM).

\begin{figure}[t]
\centering
	\includegraphics[width = \linewidth]{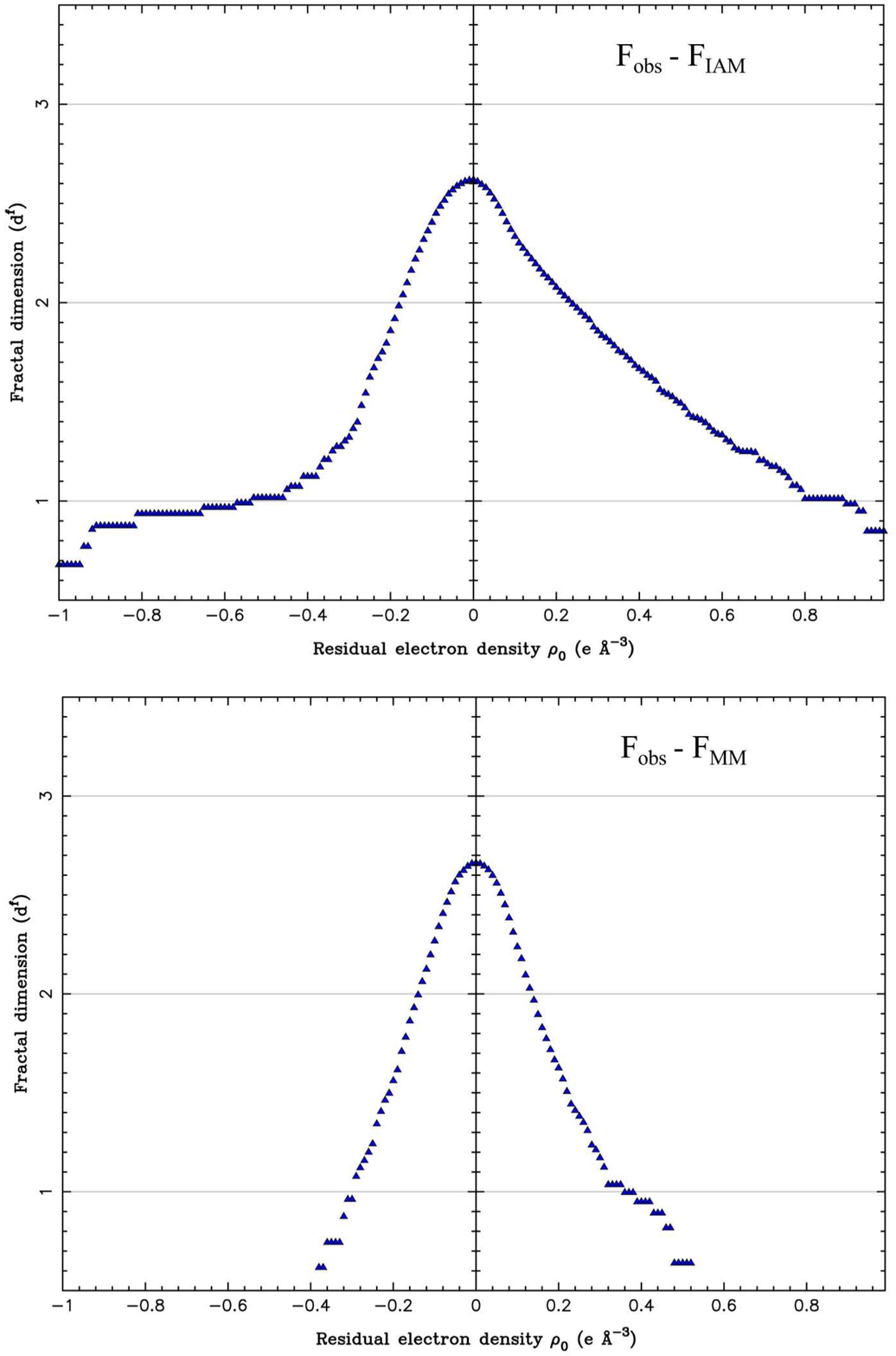}
	\caption{Fractal dimension plot of residuals for Independent Atom Model (IAM) (upper panel) and Multipole Model (MM) (lower panel) refinements of Co(NCS)$_2$(thiourea)$_2$\ X-ray data.}
	\label{fig:nicofract}
\end{figure}

A fractal dimension plot, comparing the residuals of these two methods, shown in Figure \ref{fig:nicofract} reports small and rather random discrepancies. The electron density and its Laplacian at the bond critical point have equivalent values in corresponding $M$-ligand interactions in these materials, see Table \ref{tab:nicobcp}.

Topological analysis of the experimental and theoretical electron densities are shown in Table \ref{tab:nicocharg}. Periodic Density Functional Theory (p-DFT) calculations of geometry optimizations and electron density distributions for both Co(NCS)$_2$(thiourea)$_2$\ and Ni(NCS)$_2$(thiourea)$_2$ were carried out by using CRYSTAL14 software \cite{Dovesi2014}. Optimized structural parameters and selected bond lengths are shown in Table \ref{tab:geoCoNi}. These parameters were then used to calculate values for the magnetic superexchange coupling constants with the results shown in Table \ref{tab:nicoj}. The conversion from $\Delta E$ to $J$ was done by considering a single $J$ convention in the Hamiltonian, so summations were completed over unique exchange pathways.

\begin{table}[t!]
\centering
\caption{Quantum theory of atoms in molecules integrated atomic charges for Co(NCS)$_2$(thiourea)$_2$\ and Ni(NCS)$_2$(thiourea)$_2$\ as obtained from multipolar modelling (MM) to experimental structure factor (Exp. MM) and calculated structure factor (Calc. MM), or from periodic-DFT electron density distribution (p-DFT).}
\label{tab:nicocharg}
\resizebox{\linewidth}{!}{ 
\begin{small}
\begin{tabular}{lccccc}
               & \multicolumn{3}{l}{Co(NCS)$_2$(thiourea)$_2$}                & \multicolumn{2}{l}{Ni(NCS)$_2$(thiourea)$_2$}        \\\hline\hline
Atomic charges & Exp. MM               & Calc. MM & p-DFT  & Calc. MM              & p-DFT    \\
$M$              & 1.233                 & 1.348    & 1.294  & 0.905                 & 1.154    \\
\multicolumn{6}{c}{NCS$^-$}                  \\
S(1)           & -0.304                & -0.192   & -0.140 & -0.264                & 0.135    \\
N(1)           & -0.757                & -0.785   & -1.219 & -0.583                & -1.203   \\
C(1)           & 0.058                 & 0.153    & 0.303  & 0.114                 & 0.307    \\
NCS$^-$        & -1.003                & -0.824   & -1.056 & -0.733                & -0.761   \\
\multicolumn{6}{c}{thiourea}                  \\
S(2)           & -0.110                & -0.396   & 0.144  & -0.417                & -0.087   \\
N(2)           & -1.123                & -0.877   & -1.093 & -0.916                & -1.099   \\
N(3)           & -1.153                & -0.810   & -1.102 & -0.927                & -1.095   \\
C(2)           & 0.737                 & 0.704    & 0.683  & 0.649                 & 0.685    \\
H(2A)          & 0.535                 & 0.471    & 0.454  & 0.473                 & 0.443    \\
H(2B)          & 0.517                 & 0.351    & 0.442  & 0.463                 & 0.455    \\
H(3A)          & 0.424                 & 0.345    & 0.447  & 0.483                 & 0.438    \\
H(3B)          & 0.572                 & 0.342    & 0.437  & 0.476                 & 0.446    \\
thiourea        & 0.399                 & 0.130    & 0.412  & 0.284                 & 0.186    \\\hline
\end{tabular}
\end{small}} 
\end{table}

\clearpage

\begin{table}[t]
\centering
\caption{Single crystal X-ray data and refinements details for Co(NCS)$_2$(thiourea)$_2$ and Ni(NCS)$_2$(thiourea)$_2$.}
\label{tab:xrayCoNi}
\resizebox{\columnwidth}{!}{ 
\begin{small}
\begin{tabular}{lcc}
\multicolumn{1}{l}{Compound}         & Co(NCS)$_2$(thiourea)$_2$                     & Ni(NCS)$_2$(thiourea)$_2$           \\
\hline                                                            
\hline                                                            
Formula                              & Co$_1$S$_4$N$_6$C$_4$H$_8$ & Ni$_1$S$_4$N$_6$C$_4$H$_8$ \\
M.W. / g$\cdot$mol$^{-1}$            & 327.34                     & 327.10                \\
Temperature / K                      & 100(2)                     & 173(2)          \\
Crystal system                       & primitive                  & primitive       \\
Space group                          & $P\overline{1}$                      & $P\overline{1}$           \\
a / \AA                            & 3.82030(10)                & 3.7615(2)     \\
b / \AA                            & 7.5341(2)                  & 7.5273(5)     \\
c / \AA                            & 10.0847(2)                 & 10.0744(7)     \\
$\alpha$ / $^\circ$                  & 92.065(2)                  & 92.129(5)     \\
$\beta$ / $^\circ$                   & 98.230(2)                  & 98.055(5)     \\
$\gamma$ / $^\circ$                  & 104.835(2)                 & 104.080(5)     \\
Volume / \AA$^3$                     & 276.896(2)                 & 273.19(3)     \\
Z                                    & 1                          & 1               \\
$\rho_{calc}$ / g$\cdot$cm$^{-3}$    & 1.961                      & 1.988           \\
$\mu$ / mm$^{-1}$                    & 2.275                      & 2.513           \\
2$\theta$ range /$^\circ$            & 4.1 - 90.7                 & 4.1 - 26.2      \\
$h$ index range                      & -7 / 7                     & -4 / 4        \\
$k$ index range                      & -15 / 14                   & -9 / 9        \\
$l$ index range                      & -20 / 20                   & -13 / 13        \\
Collected/unique                     & 40906 / 4662               & 4059 / 2315   \\
R$_{int}$                            & 0.040                      & 0.039           \\
R$_\sigma$                           & 0.018                      & 0.049           \\
                                     &\multicolumn{2}{c}{Spherical atom refinement}                                                                                                         \\
Data/restr./param.                   & 4527/0/71                  & 2315/0/87       \\
GooF on F$^2$                        & 1.057                      & 1.028           \\
R indexes [I$\geq$2$_\sigma$]        & R1 = 0.022                 & R1 = 0.030      \\
                                     & $w$R2 = 0.06               & $w$R2 = 0.08    \\
Residuals /e\AA$^{-3}$               & 0.75/-0.95                 & 0.64/-0.32      \\
                                     &\multicolumn{2}{c}{Multipole model refinement}                                                                             \\
Data/restr./param.                   & 4650/0/279       &         \\
GooF on F$^2$                        & 1.069            &         \\
R indexes [all data]                 & R1 = 0.019       &         \\
                                     & $w$R2 = 0.05     &         \\
Residuals / e\AA$^{-3}$              & 0.63/-0.75       &         \\
\hline           
\end{tabular}
\end{small}} 
\end{table}

\begin{table}[t]
\caption{Selected bond critical point properties for Co(NCS)$_2$(thiourea)$_2$\ and Ni(NCS)$_2$(thiourea)$_2$ where $M$ is the transition metal ion Ni(II) or Co(II). $\rho_{bcp}$ and $\bigtriangledown\rho_{bcp}$ are the electron density (\textit{e}/bohr$^3$) and its Laplacian (\textit{e}/bohr$^5$). $\epsilon$ is the dimensionless bond ellipticity.}{
\label{tab:nicobcp}
\resizebox{\linewidth}{!}{ 
\begin{tabular}{lccccccccc}
          & \multicolumn{6}{l}{Co(NCS)$_2$(thiourea)$_2$}                                                       & \multicolumn{3}{l}{Ni(NCS)$_2$(thiourea)$_2$}               \\\hline\hline
          & \multicolumn{3}{c}{Exp. MM}                              & \multicolumn{3}{c}{Calc. MM}                             & \multicolumn{3}{c}{Calc. MM}    \\
bond      & $\rho_{bcp}$ & $\bigtriangledown\rho_{bcp}$ & $\epsilon$ & $\rho_{bcp}$ & $\bigtriangledown\rho_{bcp}$ & $\epsilon$ & $\rho_{bcp}$  & $\bigtriangledown\rho_{bcp}$ & $\epsilon$ \\
$M$---N       & 0.069(2)     & 0.348(1)                     & 0.41       & 0.071        & 0.341                        & 0.14       & 0.073         & 0.343                        & 0.01       \\
$M$---S       & 0.031(1)     & 0.126(0)                     & 0.75       & 0.035        & 0.118                        & 0.01       & 0.038         & 0.116                        & 0.05       \\
$M$---S       & 0.034(1)     & 0.125(0)                     & 0.30       & 0.038        & 0.130                        & 0.02       & 0.041         & 0.126                        & 0.05       \\
$M$---$M$       & 0.016(1)     & 0.046(0)                  & \textit{1.66} & 0.016        & 0038   & \textit{1.94}     & -             & -       & -          \\
\multicolumn{10}{c}{NCS$^-$} \\
N(1)-C(1) & 0.448(6)     & -0.075(10)                   & 0.09       & 0.454        & -0.716                       & 0.01       & 0.454         & -1.009                       & 0.02       \\
C(1)-S(1) & 0.255(7)     & -0.533(6)                    & 0.10       & 0.197        & 0.064                        & 0.06       & 0.2           & -0.247                       & 0.03       \\
\multicolumn{10}{c}{thiourea} \\
S(2)-C(2) & 0.209(6)     & -0.357(4)                    & 0.25       & 0.182        & -0.249                       & 0.05       & 0.182         & -0.192                       & 0.07       \\
C(2)-N(2) & 0.366(5)     & -1.111(8)                    & 0.12       & 0.342        & -0.935                       & 0.10       & 0.349         & -0.814                       & 0.21       \\
C(2)-N(3) & 0.354(5)     & -1.116(8)                    & 0.11       & 0.342        & -0.899                       & 0.11       & 0.351         & -0.839                       & 0.23       \\\hline
\end{tabular}
}}
\end{table}

\begin{table}[b]
\centering
\caption{Peridoc-DFT geometry optimizations of Co(NCS)$_2$(thiourea)$_2$\ and Ni(NCS)$_2$(thiourea)$_2$ where $M$ represents either Ni(II) or Co(II).}{
\label{tab:geoCoNi}
\begin{small}
\begin{tabular}{lcc}
\multicolumn{1}{l}{Compound}   & Co(NCS)$_2$(thiourea)$_2$      & Ni(NCS)$_2$(thiourea)$_2$     \\\hline\hline
Space group                    & $P\overline{1}$  &$P\overline{1}$      \\
a / \AA                      & 3.88868     & 3.83801    \\
b / \AA                      & 7.83699     & 7.60047    \\
c / \AA                      & 10.22564    & 10.18991   \\
$\alpha$ / $^\circ$            & 96.04319    & 92.10032   \\
$\beta$ / $^\circ$             & 98.42657    & 98.47791   \\
$\gamma$ / $^\circ$            & 108.42653   & 99.92120   \\
Volume / \AA$^3$               & 291.884     & 289.014    \\
               \multicolumn{3}{c}{Bond lengths and angles} \\
$M$---S / \AA                    &  2.567      &  2.549   \\
$M$---S / \AA                    &  2.605      &  2.585   \\
$M$---N / \AA                    &  2.052      &  2.026   \\
$M$---$M$ / \AA                    &  3.889      &  3.838   \\
$M$---S---$M$ / $^\circ$               &  97.51      &  96.77   \\
S---$M$---S / $^\circ$               &  82.49      &  83.23   \\\hline           
\end{tabular}
\end{small}} 
\end{table}  
\end{sloppypar}
\clearpage
\balance
\bibliography{2SI_bib}